\documentclass[traditabstract,10pt]{aa}
\usepackage[T1]{fontenc}
\usepackage[english]{babel}
\usepackage{amsmath, amssymb, amsfonts, latexsym}
\usepackage{epstopdf}
\usepackage[breaklinks, colorlinks, linkcolor=blue, citecolor=blue]{hyperref} 
\usepackage{ifthen}
\usepackage[usenames,dvipsnames,table]{xcolor}
\usepackage{fixltx2e}
\usepackage{placeins}
\usepackage{graphicx}
\usepackage{float}
\usepackage[nonamebreak]{natbib}
\bibpunct{(}{)}{;}{a}{}{,}
\providecommand{\sorthelp}[1]{}

\def\planck{\textsc{Planck}}

\newcommand{\Acib}{\ensuremath{\mathrm{A_\text{cib}}}}
\newcommand{\Asz}{\ensuremath{\mathrm{A_\text{SZ}}}}
\newcommand{\Aszcib}{\ensuremath{\mathrm{A_\text{SZxCIB}}}}
\newcommand{\Aksz}{\ensuremath{\mathrm{A_\text{kSZ}}}}
\newcommand{\Plik}{{\tt Plik}}
\newcommand{\PlikTT}{{\tt PlikTT}}
\newcommand{\plikALL}{{\tt PlikALL}}
\newcommand{\SPT}{{\tt{SPT}}}
\newcommand{\ACT}{{\tt{ACT}}}
\newcommand{\spthigh}{{\tt SPT\_high}}
\newcommand{\sptlow}{{\tt SPT\_low}}
\newcommand{\VHL}{\textsc{VHL}}
\newcommand{\SN}{\textsc{SNI}a}
\newcommand{\SNplot}{\SN}
\newcommand{\hlp}{{\tt{HiLLiPOP}}}
\newcommand{\hillipop}{{\hlp}}
\newcommand{\CAMEL}{\textsc{CAMEL}}
\newcommand{\hlpTT}{{\tt hlpTT}}

\newcommand{\hlpALL}{{\tt hlpALL}}
\newcommand{\hlpALLps}{{\tt hlpALLps}}
\newcommand{\hlpTE}{{\tt hlpTE}}
\newcommand{\hlpPS}{{\tt hlpTTps}}

\newcommand{\lollipop}{{\tt{Lollipop}}}
\newcommand{\Commander}{{\tt{Commander}}}

\newcommand{\mnu}{\ensuremath{\mathrm{\Sigma m_{\nu}}}}

\newcommand{\lambdaCDM}{\ensuremath{\mathrm{\Lambda{CDM}}}}
\newcommand{\MnulambdaCDM}{$\nu\lambdaCDM$}
\newcommand{\baselambdaCDM}{\ensuremath{\mathrm{\Lambda{CDM}(1\nu)}}}
\newcommand{\MnubaselambdaCDM}{$\ensuremath{\mathrm{\nu\Lambda{CDM}(1\nu)}}$}

\newcommand{\troisnulambdaCDM}{\ensuremath{\mathrm{\Lambda{CDM}(3\nu)}}}
\newcommand{\MnutroisnulambdaCDM}{$\ensuremath{\mathrm{\nu\Lambda{CDM}(3\nu)}}$}

\newcommand{\alens}{\ensuremath{\mathrm{A_{L}}}}

\newcommand{\p}{\ensuremath{\mathrm{p}}}

\newcommand{\mead}{\textsc{Mead}}
\newcommand{\taka}{\textsc{Takahashi}}

\newcommand{\plik}{{\tt Plik}}

\newcommand{\mksym}[1]{\ifmmode {\rm #1}\else #1\fi}
\newcommand{\dataplus}{{+}}

\newcommand{\BAO}{\mksym{BAO}}

\newcommand{\TT}{\mksym{TT}}

\newcommand{\planckTTonly}{\planck\ \TT}

\newcommand{\lowTEB}{\mksym{lowTEB}}

\newcommand{\planckTT}{\planckTTonly\dataplus\lowTEB}

\newcommand{\taureio}{\ensuremath{\tau_{\rm reio}}}
\newcommand{\As}{\ensuremath{\mathrm{A_{\rm s}}}}

\newcommand{\ns}{\ensuremath{n_{\rm s}}}

\newcommand{\lcdm}{{$\rm{\Lambda CDM}$}}

\newcommand{\Alens}{\ensuremath{A_{\rm L}}}

\providecommand{\planck}{\Planck}

\providecommand{\text}[1]{\rm{#1}}

\providecommand{\muK}{\mu\rm{K}}

\newcommand{\eV}{\,\text{eV}}

\providecommand{\CAMB}{{\tt CAMB}}

\providecommand{\CLASS}{{\tt CLASS}}

\newcommand{\begm}{\begin{pmatrix}}
\newcommand{\enm}{\end{pmatrix}}

\newcommand\ba{\begin{eqnarray}}
\newcommand\ea{\end{eqnarray}}
\newcommand\bea{\begin{eqnarray}}
\newcommand\eea{\end{eqnarray}}

\newcommand\be{\begin{equation}}
\newcommand\ee{\end{equation}}

\def\pmb#1{\setbox0=\hbox{#1}%
    \kern-.025em\copy0\kern-\wd0
    \kern.05em\copy0\kern-\wd0
    \kern-.025em\raise.0433em\box0}

\def\p2Y{\;_2Y}
\def\m2Y{\;_{-2}Y}
\def\beglet{
  \addtocounter{equation}{1}%
  \setcounter{parentequation}{\value{equation}}%
  \setcounter{equation}{0}%
  \def\theequation{\arabic{parentequation}\alph{equation}}%
  \ignorespaces
}
\def\endlet{
  \setcounter{equation}{\value{parentequation}}%
  \def\theequation{\arabic{equation}}%
}
\providecommand{\beglet}{\begin{subequations}}
\providecommand{\endlet}{\end{subequations}}

\def\setsymbol#1#2{\expandafter\def\csname #1\endcsname{#2}}
\def\getsymbol#1{\csname #1\endcsname}

\newbox\tablebox    \newdimen\tablewidth
\def\leaderfil{\leaders\hbox to 5pt{\hss.\hss}\hfil}
\def\tablenote#1 #2\par{\begingroup \parindent=0.8em
    \abovedisplayshortskip=0pt\belowdisplayshortskip=0pt
    \noindent
    $$\hss\vbox{\hsize\tablewidth \hangindent=\parindent \hangafter=1 \noindent
    \hbox to \parindent{$^#1$\hss}\strut#2\strut\par}\hss$$
    \endgroup}

\def\L2{\ifmmode L_2\else $L_2$\fi}

\def\DeltaT{\ifmmode \Delta T\else $\Delta T$\fi}
\def\deltat{\ifmmode \Delta t\else $\Delta t$\fi}
\def\fknee{\ifmmode f_{\rm knee}\else $f_{\rm knee}$\fi}
\def\Fmax{\ifmmode F_{\rm max}\else $F_{\rm max}$\fi}
\def\solar{\ifmmode{\rm M}_{\mathord\odot}\else${\rm M}_{\mathord\odot}$\fi}
\def\Msolar{\ifmmode{\rm M}_{\mathord\odot}\else${\rm M}_{\mathord\odot}$\fi}
\def\Lsolar{\ifmmode{\rm L}_{\mathord\odot}\else${\rm L}_{\mathord\odot}$\fi}
\def\inv{\ifmmode^{-1}\else$^{-1}$\fi}
\def\mo{\ifmmode^{-1}\else$^{-1}$\fi}
\def\sup#1{\ifmmode ^{\rm #1}\else $^{\rm #1}$\fi}
\def\expo#1{\ifmmode \times 10^{#1}\else $\times 10^{#1}$\fi}
\def\,{\thinspace}
\def\lsim{\mathrel{\raise .4ex\hbox{\rlap{$<$}\lower 1.2ex\hbox{$\sim$}}}}
\def\gsim{\mathrel{\raise .4ex\hbox{\rlap{$>$}\lower 1.2ex\hbox{$\sim$}}}}

\def\simprop{\mathrel{\raise .4ex\hbox{\rlap{$\propto$}\lower 1.2ex\hbox{$\sim$}}}}
\def\deg{\ifmmode^\circ\else$^\circ$\fi}
\def\pdeg{\ifmmode $\setbox0=\hbox{$^{\circ}$}\rlap{\hskip.11\wd0 .}$^{\circ}
          \else \setbox0=\hbox{$^{\circ}$}\rlap{\hskip.11\wd0 .}$^{\circ}$\fi}
\def\arcs{\ifmmode {^{\scriptstyle\prime\prime}}
          \else $^{\scriptstyle\prime\prime}$\fi}
\def\arcm{\ifmmode {^{\scriptstyle\prime}}
          \else $^{\scriptstyle\prime}$\fi}
\newdimen\sa  \newdimen\sb
\def\parcs{\sa=.07em \sb=.03em
     \ifmmode \hbox{\rlap{.}}^{\scriptstyle\prime\kern -\sb\prime}\hbox{\kern -\sa}
     \else \rlap{.}$^{\scriptstyle\prime\kern -\sb\prime}$\kern -\sa\fi}
\def\parcm{\sa=.08em \sb=.03em
     \ifmmode \hbox{\rlap{.}\kern\sa}^{\scriptstyle\prime}\hbox{\kern-\sb}
     \else \rlap{.}\kern\sa$^{\scriptstyle\prime}$\kern-\sb\fi}
\def\ra[#1 #2 #3.#4]{#1\sup{h}#2\sup{m}#3\sup{s}\llap.#4}
\def\dec[#1 #2 #3.#4]{#1\deg#2\arcm#3\arcs\llap.#4}
\def\deco[#1 #2 #3]{#1\deg#2\arcm#3\arcs}
\def\rra[#1 #2]{#1\sup{h}#2\sup{m}}

\def\dots{\relax\ifmmode \ldots\else $\ldots$\fi}
\def\WHzsr{\ifmmode $W\,Hz\mo\,sr\mo$\else W\,Hz\mo\,sr\mo\fi}
\def\mHz{\ifmmode $\,mHz$\else \,mHz\fi}
\def\GHz{\ifmmode $\,GHz$\else \,GHz\fi}
\def\mKs{\ifmmode $\,mK\,s$^{1/2}\else \,mK\,s$^{1/2}$\fi}
\def\muKs{\ifmmode \,\mu$K\,s$^{1/2}\else \,$\mu$K\,s$^{1/2}$\fi}
\def\muKRJs{\ifmmode \,\mu$K$_{\rm RJ}$\,s$^{1/2}\else \,$\mu$K$_{\rm RJ}$\,s$^{1/2}$\fi}
\def\muKHz{\ifmmode \,\mu$K\,Hz$^{-1/2}\else \,$\mu$K\,Hz$^{-1/2}$\fi}
\def\MJysr{\ifmmode \,$MJy\,sr\mo$\else \,MJy\,sr\mo\fi}
\def\MJysrmK{\ifmmode \,$MJy\,sr\mo$\,mK$_{\rm CMB}\mo\else \,MJy\,sr\mo\,mK$_{\rm CMB}\mo$\fi}
\def\microns{\ifmmode \,\mu$m$\else \,$\mu$m\fi}

\def\muK{\ifmmode \,\mu$K$\else \,$\mu$\hbox{K}\fi}
\def\microK{\ifmmode \,\mu$K$\else \,$\mu$\hbox{K}\fi}
\def\muW{\ifmmode \,\mu$W$\else \,$\mu$\hbox{W}\fi}
\def\kms{\ifmmode $\,km\,s$^{-1}\else \,km\,s$^{-1}$\fi}
\def\kmsMpc{\ifmmode $\,\kms\,Mpc\mo$\else \,\kms\,Mpc\mo\fi}
\providecommand{\sorthelp}[1]{}

\begin{document}

%%%%%%%%%%%%%%%%%%%%%%%%%%%%%%%%%%%%%%%%%%%%%%%%%%%%%%%%%%%%%%%%%%%%%%%%%%
% HEADER
%%%%%%%%%%%%%%%%%%%%%%%%%%%%%%%%%%%%%%%%%%%%%%%%%%%%%%%%%%%%%%%%%%%%%%%%%%

\title{Cosmological constraints on the neutrino mass including systematic uncertainties}
\author{F.~Couchot\inst{1}, S.~Henrot-Versill\'e\thanks{Corresponding author: \href{mailto:versille@lal.in2p3.fr}{versille@lal.in2p3.fr}}\inst{1}, O.~Perdereau\inst{1}, S.~Plaszczynski\inst{1},\\ B.~Rouill\'e d'Orfeuil\inst{1}, M.~Spinelli\inst{1,2}, and M.~Tristram\inst{1}}
%\thanks{Corresponding author: \href{mailto:}{}}}
\authorrunning{F. Couchot et al.}
\institute{Laboratoire de l'Acc\'el\'erateur Lin\'eaire, Univ. Paris-Sud, CNRS/IN2P3, Universit\'e Paris-Saclay, Orsay, France
\and 
Department of Physics and Astronomy, University of the Western Cape, Robert Sobukwe Road, Bellville 7535, South Africa}

\abstract{When combining cosmological and oscillations results to constrain the neutrino sector, the 
question of the  propagation of systematic uncertainties is often raised. 
We address this issue in the context of the derivation of an upper bound
on the sum of the neutrino masses (\mnu) 
with recent cosmological data. This work is performed within the \lambdaCDM\ model extended to \mnu, for which
we advocate the use of three mass-degenerate neutrinos. We focus on the study of systematic uncertainties linked to 
the foregrounds modelling in CMB data analysis,
and on the impact of the present knowledge of the reionisation optical depth. 
This is done
through the use of different likelihoods built from \planck\ data.
Limits on \mnu\ are derived with various combinations of data, including the latest Baryon Acoustic Oscillations (\BAO) and
Type Ia Supernovae (\SN) results. 
We also discuss  the impact of the preference for current CMB data for
amplitudes of the gravitational lensing distortions higher than expected within the \lambdaCDM\ model,
and add the \planck\ CMB lensing.
We then derive a robust upper limit:
$\mnu<\ 0.17\hbox{\ eV\ at\ }95\%\ \hbox{CL}$, including 0.01\eV\ of foreground systematics.
We also discuss the neutrino mass repartition and show that today's data do not
allow one  to disentangle normal from inverted hierarchy. The impact on the other 
cosmological
parameters is also reported, 
for different assumptions on the 
neutrino mass repartition, and different
high and low multipole CMB likelihoods.
}
%TODO - ell --ell

\date{\today}

\authorrunning{F. Couchot et al.}
%\titlerunning{Robust limit on \mnu\ with cosmological data}

\maketitle

%\tableofcontents
%\vskip 1cm

%%%%%%%%%%%%%%%%%%%%%%%%%%%%%%%%%%%%%%%%%%%%%%%%%%%%%%%%%%%%%%%%%%%%%%%%%%
%INTRODUCTION
%%%%%%%%%%%%%%%%%%%%%%%%%%%%%%%%%%%%%%%%%%%%%%%%%%%%%%%%%%%%%%%%%%%%%%%%%%
\section*{Introduction}

In the last decade, cosmology has entered a precision era, confirming  
the six parameters \lambdaCDM\ concordance model 
with unprecedented accuracy. 
This allows us to open the parameters'
space, and to  
confront the corresponding extensions 
with data. 
In the following, we explore the neutrino sector.
We only deal with three standard neutrinos species~\citep{ALEPH:2005ab},
and 
focus on the 
extension to the sum of the neutrino masses (\mnu).
Moreover,  
the neutrino mass splitting scenario  
has been set up to  match the neutrino oscillation
results. A three mass-degenerate neutrinos  model
is advocated for and used throughout this study. 
It must be noted that 
the assumptions on the neutrino mass scenario
have already been shown to
be of particular importance 
for the derivations  of cosmological 
results~\citep[for example in][]{2011MNRAS.418..346M}. 
 
Recent works~\citep[for instance][]{Alam:2016hwk, Sherwin:2016tyf, Giusarma:2016phn,Yeche:2017upn,Vagnozzi:2017ovm}
on the derivation of upper bounds on \mnu\,
usually take the  
Cosmological Microwave Background (CMB) likelihoods as granted.
Furthermore, no uncertainty from the
analysis of this 
cosmological probe  is propagated until the final results.
In this paper, we
investigate the systematic uncertainties linked to the modelling of foreground residuals
in the \planck\ CMB likelihood 
implementations.

To address this issue, the most accurate method would have been
to make use of full end to end simulations, including an exhaustive description
of the foregrounds. This is not possible given the
actual knowledge of the foreground's physical properties.
Instead, we propose a comparison of the results derived from 
different likelihoods built 
from the \planck\ 2015 data release, and 
based on different foreground assumptions. Namely  
the public \plik\ and the \hlp\ likelihoods are examined for the high-$\ell$ part.
We also investigate the impact of our current knowledge on the reionisation optical depth ($\taureio$).
For the low-$\ell$ part, the \lowTEB\ likelihood is compared
to the combination of the \Commander\ likelihood with an auxiliary constraint on the $\taureio$ parameter, derived
from the last \planck\ 2016 measurements~\citep{planck2014-a25}.

The differences of the impact of the foreground modellings are twofold:     
On one hand they show up as slight deviations on the \mnu\ bounds inferred from the different
likelihoods, and, on the other hand, they manifest themselves in the form
of different values of the amplitude of the gravitational lensing distortions 
(\alens). Indeed, fitting for \alens\ represents 
a direct test of the accuracy and robustness of the likelihood with respect to the \lambdaCDM\ model~\citep{couchot:2015}. 
We also address this point, and discuss how it is linked to \mnu.

Derivations of systematic uncertainties on \mnu\ are performed for different combinations
of cosmological data: The \planck\ temperature and polarisation 
likelihoods, 
the latest \BAO\ data from Boss DR12, and \SN,
as well as the direct measurement of the
lensing distortion field power spectrum from \planck. 

We also address the question of the sensitivity of the combination of those datasets
to the neutrinos mass hierarchy. 

We start with a description of the standard cosmology, the impact of massive
neutrinos, and their mass repartition,
as well as the profile likelihood method (Sect.~\ref{designations}).
In Sect.~\ref{datasets}, we describe the likelihoods and datasets. 
Turning to the \mnu\ constraints, we first focus on the results obtained
with CMB temperature data for different 
likelihoods at intermediate multipoles.  
We investigate different choices for the low-$\ell$ likelihoods, 
and examine the pros and cons of the use of high-angular-resolution datasets.
In Sect.~\ref{SNBAO}, we derive the \mnu\ constraints obtained when combining
CMB temperature, \BAO\ and \SN\ data, and check the robustness of the results with respect to
the high-$\ell$ likelihoods. 
The choices for the low-$\ell$ parts are compared. 
A cross-check of the results is performed using the temperature-polarisation TE correlations.
Then, the impact of the observed tension on \alens\ is further discussed,
followed by the combination of the data with the CMB lensing.
The neutrino mass hierarchy question is addressed in this context.
In Sect.~\ref{sec:mnu_All},  we discuss the (TT+TE+EE)  combination with BAO and SN data,
with and without CMB lensing.
Finally, we derive the cosmological parameters
and illustrate their variations depending on the assumptions on the neutrino mass repartition,
the low-$\ell$ likelihoods, and the fact that we release or do not release \mnu\
in the fits. 

\section{Phenomenology and methodology}
\label{designations}
This section discusses the standard cosmology and 
the role of neutrinos in the Universe's thermal history.
We then briefly review the current constraints coming from the observation
of the neutrino oscillations phenomenon,
and discuss the mass hierarchy. 
A definition of the $\lambdaCDM$ models 
considered for this paper
is given.
The statistical
methodology based on profile likelihoods
is also presented.

\subsection{Standard Cosmology}
\label{sec:lcdm}

The ``standard'' cosmological model describes the evolution of a homogeneous and isotropic Universe, the geometry 
of which is given by the Friedman-Robertson-Walker metric, following General Relativity. In this framework, the theory reduces to the well-known Friedman equations. The Universe is assumed to be filled with several components, of different nature and evolution (matter, radiation,...). Their inhomogeneities are accounted for as small perturbations of the metric. 
In the \lambdaCDM\ model, the Universe's geometry is assumed to be euclidean (no curvature) and its constituents are dominated today by 
a cosmological constant  ($\Lambda$), associated with dark energy,
and cold dark matter; it also includes radiation, baryonic matter and three neutrinos.
Density anisotropies are assumed to result from the evolution of primordial power spectra, 
and only purely adiabatic scalar modes are assumed. 

The minimal \lambdaCDM\ model is described with only six parameters. Two of them
describe the primordial scalar 
mode power spectrum: the amplitude (\As), and the spectral index (\ns).
Two other parameters represent the reduced energy densities today:
$\omega_{\rm b}=\Omega_{\rm b}h^2$, for the baryon, 
and $\omega_{\rm c}=\Omega_{\rm c}h^2$ for the cold dark matter.
The last two parameters are the angular size of the sound horizon at decoupling, $\theta_S $,
and the reionisation optical depth (\taureio).
In this chosen parameterisation, 
$H_0$ is derived in a non-trivial way from the above parameters.
In addition, the sum of the neutrino masses is usually fixed to $\mnu=0.06$~\eV\ based on oscillation 
constraints~\citep{Forero:2012,Forero:2014bxa,Capozzi:2016rtj}: This is discussed in Sect.~\ref{sec:nu_oscillations}.

Departures from the \lambdaCDM\ model assumptions are often studied by extending its parameter space and testing
it against the data, for instance, through the inclusion of $\Omega_k$ for non-euclidean geometry, $N_{\hbox{\tiny{eff}}}$
for the number of effective relativistic species, or $Y_p$ for 
the primordial mass fraction of $^{4}\hbox{He}$ during BBN. 
In addition to those physics-related parameters, a phenomenological parameter, \alens, has been introduced~\citep{Alens} to scale the 
deflection power spectrum which is used to lens the primordial CMB power spectra. 
This parameter permits to size the (dis-)agreement of the data with the \lambdaCDM\ lensing distortion predictions.
Testing that its value, inferred from data, is compatible with one is a thorough consistency check \citep[we refer to e.g.][]{Calabrese:2008rt,planck2014-a15,couchot:2015}. 
In this work, we use the \alens\ consistency check in the context of the constraints on \mnu. In practice, it means that we check the value of \alens\ (using \lambdaCDM+\alens\ model) for each dataset on which we  then report a \mnu\ limit (using \MnulambdaCDM\ model, i.e. with $\alens=1$).

\subsection{Neutrinos in cosmology} 
\label{sec:nu_cosmo}

One of the generic features of the standard hot big bang model is the existence of 
a relic cosmic neutrino background.
In parallel, the observation of the neutrino oscillation phenomena requires that those particles are massive,
and  establishes the existence of flavour mixed-mass eigenstates 
(cf. Sect.~\ref{sec:nu_oscillations})~\citep{Pontecorvo:1957cp,Maki:1962mu}.
As far as cosmology is concerned, depending on the mass of the lightest neutrino~\citep{Bilenky:2001rz},
this implies that there are at least two non-relativistic species today.
Massive neutrinos therefore impact the energy densities of the Universe and its evolution.

Initially neutrinos are coupled  to  the  primeval
plasma. As the Universe cools down, they decouple
from the rest of the plasma at a temperature up to a few MeV depending on their flavour~\citep{Dolgov:2002wy}.
This decoupling is fairly well approximated as an instantaneous process~\citep{kolb1994early,Dodelson:1282338}.
Given the fact that, with today's observational constraints,
neutrinos can be considered as relativistic 
at recombination~\citep{Lesgourgues:2006nd}. 
In addition, for $\mathrm{m_{\nu}}$ in the range from $10^{-3}$ to 1~\eV, they should be
counted as radiation at the matter-radiation equality redshift,  $z_{\hbox{\tiny{eq}}}$,
and as non-relativistic matter today~\citep{jl_neutrino_and_planck,lesgourgues}, which is measured through $\Omega_m$.
\mnu\ is therefore correlated to both $z_{\hbox{\tiny{eq}}}$ and $\Omega_m$.

The induced modified background evolution is reflected in the relative position and amplitude of the peaks 
of the CMB power spectra (through
$z_{\hbox{\tiny{eq}}}$). It also affects the CMB  anisotropies power spectra 
at intermediate or high multipole ($\ell\gtrsim 200$) as potential shifts of the power spectrum 
due to a change in the angular distance of the sound horizon at decoupling. Finally it also leaves an imprint on
the slope of the low-$\ell$ tail due to the late Integrated Sachs Wolfe (ISW) effect.
An additional effect of massive neutrinos comes from the fact that they affect
the photon temperature through the early ISW effect. As a result a reduction of the
CMB temperature power spectrum below $\lesssim 500$ is observed. 

On the matter power spectrum side, two effects are induced by the massive neutrinos.
In the early Universe, they free-stream out of potential wells, damping matter perturbations on scales 
smaller than the horizon at the non-relativistic transition. This results in a suppression of the $\hbox{P(k)}$ 
at large $\hbox{k}$ which also depends on the individual masses repartition~\citep{Hu:1997mj,Lesgourgues:2006nd}. 
At late time, the non-relativistic neutrino masses modify the matter density, which tends
to slow down the clustering. 

CMB anisotropies are lensed by large-scale structures (LSS).
Measuring CMB gravitational lensing therefore provides a constraint on the matter power spectrum
on scales where the effects of massive neutrinos are small but still sizeable~\citep{Kaplinghat:2003bh,Lesgourgues:2005yv}.

\subsection{Neutrino mass hierarchy}
\label{sec:nu_oscillations}

As stated above, we have to choose a neutrino mass splitting scenario to define the \lambdaCDM\ model.
In general, CMB data analyseis that aim at
measuring cosmological parameters not related to 
the neutrino sector
\citep[including \planck\ papers, e.g. ][]{planck2014-a15} 
are done assuming  two massless neutrinos and one massive neutrino, while fixing $\mnu=0.06$~\eV. 

For the work of this paper, our choice is motivated considering neutrino oscillation data.
More precisely, we use the differences of squared neutrino masses deduced from 
the best fit values of the global 3$\nu$ oscillation analysis based on the work of \citet{Capozzi:2016rtj}:

\begin{eqnarray}
        \label{eq:dm12}
        \Delta m^2_{21}=m_2^2-m_1^2 &=& \ 7.37\ 10^{-5}\ \mathrm{eV}^2 \\
        \label{eq:dm13}
        \Delta m^2 = m_3^2-(m_1^2+m_2^2)/2. &=& +2.50\ 10^{-3}\ \mathrm{eV}^2\ \hbox{(NH)}\\
        \label{eq:dm13ih}
        &=& -2.46\ 10^{-3}\ \mathrm{eV}^2\ \hbox{(IH)}
,\end{eqnarray}
where the two usual scenarios are considered: The normal  (NH) and the inverted hierarchy (IH),
for which the lightest neutrino is the one of the first and third generation respectively.

\begin{figure}
        \centering
        \includegraphics[width=.49\textwidth]{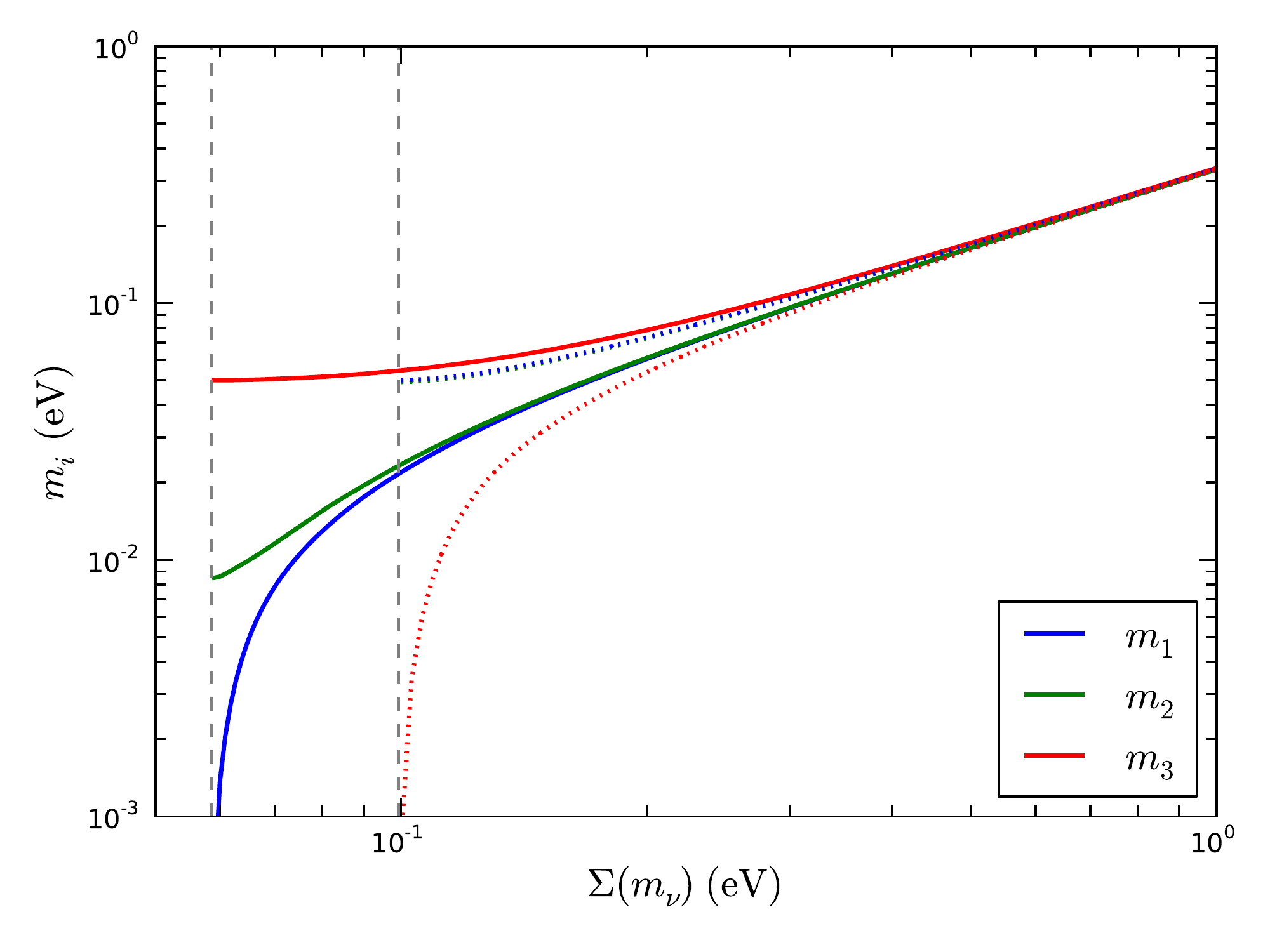}
        \caption{Individual neutrino masses as a function of  \mnu\ for the two hierarchies  (NH : plain line, IH dotted lines), under the assumptions given by equations \ref{eq:dm12} and \ref{eq:dm13}-\ref{eq:dm13ih}. The vertical dahed lines outline the minimal \mnu\ value allowed in each case (corresponding to one massless neutrino generation). The log vertical axis prevents from the difference between $m_1$ 
and $m_2$ to be resolved in IH. }
        \label{fig:mnu_mass_splitting}
\end{figure}

Individual masses can be computed numerically under the above assumptions, for each mass hierarchy, as a function of \mnu\ , as highlighted in Figure~\ref{fig:mnu_mass_splitting}
(see also~\cite{jl_neutrino_and_planck}). In each hierarchy, Equations \ref{eq:dm12}-\ref{eq:dm13ih} impose a lower bound on \mnu\, , corresponding to the case where the lightest mass is strictly null (numerically, $\sim$0.059 and $\sim$0.099 eV for NH and IH, respectively); also shown in Figure~\ref{fig:mnu_mass_splitting} as vertical dashed lines. 

Those results show that, given the oscillation constraints, neutrino masses are nearly degenerate for $\mnu\gtrsim 0.25$~\eV.
Moreover, given the current cosmological probes (essentially CMB and \BAO\ data), we observe almost no difference in \mnu\ constraints when comparing 
results obtained with one of the two hierarchies with the case with three mass-degenerate 
neutrinos for which the mass repartition is such that each neutrino carries $\mnu/3.$~\citep[we refer to Sect.~\ref{sec:mnu_nhorih} and][]{Giusarma:2016phn, Vagnozzi:2017ovm, Schwetz:2017fey}. Indeed, as shown in~\citet{palanque:2015}, the difference is less than 0.3\% in the 3D linear matter power spectrum and is reduced even to less than 0.05\% when considering the 1D flux power spectrum
~\citep[see also][]{2011MNRAS.410.1647A}.
This justifies the simplifying choice of the three mass-degenerate neutrinos scenario, which is used in this paper.

In Sect.~\ref{sec:mnu_nhorih}, we show
that this is not equivalent to the configuration where the total mass is entirely given to one massive neutrino with the two other neutrinos being 
massless.

\subsection{Constraints on \mnu\ and degeneracies}
\label{sec:nu_degeneracies}

The inference from CMB data of a limit on \mnu in the \lambdaCDM\ framework is not trivial because of degeneracies between parameters. 
Indeed, the impact of \mnu\ on the CMB temperature power spectrum is partly degenerated with that of some of the six other parameters. 

In particular, the impact of neutrino masses on the angular-diameter distance to last scattering surface is degenerated with $\Omega_\Lambda$ (and consequently with the derived parameters $H_0$ and $\sigma_8$) in flat models and with $\Omega_k$ otherwise~\citep{hou:2014}. 
Late-time geometric measurements help in reducing this geometric degeneracy. Indeed, at fixed $\theta_S$, the BAO distance parameter $D_V(z)$ increases with increasing neutrino mass while the Hubble parameter decreases.

Another example is the correlation of \mnu\ with \As~\citep{Allison:2015qca}. As explained in Sect~\ref{sec:nu_cosmo}, $\mnu$ can impact the amplitude of the matter power spectrum and thus is directly correlated to \As\ and consequently with \taureio\ through the amplitude of the first acoustic peak (which scales like $\As e^{-2\taureio}$). 
The constraint on \mnu\ therefore depends on the low-$\ell$ polarisation likelihood, which drives the constraints on \taureio.
The addition of lensing distortions, the amplitude of which is proportional to \As, helps to break this degeneracy.

Moreover, the suppression of the small-scale power in LSS due to massive neutrinos, which imprints 
on the CMB lensing spectra, can be compensated for by an increase of the cold-dark-matter density, shifting the matter-radiation equality to early times~\citep{Hall:2012,Pan:2014}. This induces an anti-correlation between \mnu\ and $\Omega_{\rm cdm}$ when using CMB observable. On the contrary, both parameters similarly affect 
the angular diameter distance so that \BAO\ can help to break this degeneracy.

\subsection{Cosmological model}
\label{sec:boltz}

As discussed in the previous sections, the neutrino mass repartition can have significant impact on the constraints for \mnu.
By \baselambdaCDM, we  refer to the definition used in~\cite{planck2014-a15}; it assumes two massless and one massive neutrinos. 

However, in the following, we adopt a scenario with three mass-degenerate neutrinos, that is, where the neutrino generations equally share the mass (\mnu/3).
We note that this is also the model adopted in~\cite{planck2014-a15} when \mnu\ constraints have been extracted.
We also stick to this scenario when fixing \mnu\ to 0.06~\eV\ and we refer to it as \troisnulambdaCDM.

The notations \MnubaselambdaCDM\ (resp. \MnutroisnulambdaCDM) will be used to 
differentiate the case where we open the parameters' space to \mnu\
from the \baselambdaCDM\ (resp. \troisnulambdaCDM) case.

To derive the values for the observables from the cosmological model, we make use of the \CLASS\ Boltzmann solver~\citep{2011JCAP...07..034B}. 
Within this software, the non-linear effects on the matter power spectrum evolution can be included using the halofit model recalibrated as proposed in~\cite{2012ApJ...761..152T} and extended to massive neutrinos as described in~\cite{2012MNRAS.420.2551B}. 
Our baseline setup for the \mnu\ studies is to use \CLASS, including non-linear effects, tuned to a high-precision setting.

In order to compare order of magnitudes in the non-linear effects propagation, we have also used \CAMB~\citep{camb}, in which both the \taka\ and the  \mead~\citep{2016MNRAS.459.1468M} models are made available.

\subsection{Profile likelihoods}
\label{sec::prof}
The results described below were obtained from profile likelihood analyses performed with the \CAMEL\ software\footnote{\url{camel.in2p3.fr}}\citep{Henrot-Versille:2016htt}. 
As described in \cite{planck2013-XVI}, this method aims at measuring a 
parameter $\theta$ through the maximisation of the likelihood function $\cal{L}(\theta,\mu)$,
where  $\mu$  is the full set of cosmological and nuisance parameters excluding $\theta$.
For different, fixed $\theta_i$ values, a multidimensional
minimisation of the 
$\chi^2(\theta_i,\mu) = - 2 \ln{\cal{L}}(\theta_i,\mu) $ function is performed. 
The absolute minimum, $\chi^2_{min}$, of the resulting $\chi^2_{min}(\theta_i)$  
curve is by construction the (invariant) global minimum of the problem, that is, the `best fit'.
From the $\chi^2_{min}(\theta_i)-\chi^2_{min}$ curve, the so-called profile likelihood, one can derive an estimate of $\theta$ and 
its associated uncertainty~\citep{jamesbook}. 
All minimisations have been performed using the {\tt MINUIT} 
software~\citep{James:1994vla}. 
In the \mnu\ studies presented below, $95\%$ CL upper limits are derived 
following the Gaussian prescription proposed by~\cite{FC} (hereafter denoted F.C.),
as described in ~\cite{planck2013-XVI}. 

Unless otherwise explicitly stated, we use the frequentist methodology throughout this
paper. A comparison with the Bayesian approach has already been
presented in~\cite{planck2013-XVI} and~\cite{planck2014-a13}, showing that results do not depend on the
chosen statistical method for the \lambdaCDM\ model, as well as for \MnulambdaCDM.

\section{Likelihoods and datasets}
\label{datasets}
In this Section, we detail the likelihoods that are used hereafter for the derivation of the results on \mnu. 
They are summarised in Table~\ref{tab:acro} together with their related acronyms.
\begin{table*}
\centering          
\begin{tabular}{l l } 
\hline\hline       
Acronym & Description  \\
\hline                    
$\hbox{\hlpTT }$&  high-$\ell$ \hillipop\ temperature \planck\  likelihood (cf. Sect~\ref{lik_intro})\\ 
$\hbox{\hlpPS}$&   high-$\ell$  \hillipop\ temperature \planck\ likelihood with an astrophysical model of point sources\\ 
$\hbox{\PlikTT} $& public high-$\ell$ temperature \planck\ likelihood  \\ 
\hline
$\hbox{TT }$& refers to the temperature CMB data  \\ 
$\hbox{TE }$& refers to the TE CMB correlations  \\ 
$\hbox{ALL}$& refers to the combination of temperature and polarisation CMB data (incl. TT and TE)\\
\hline
$\hbox{Comm} $& \Commander\  low-$\ell$ temperature \planck\ public likelihood (cf. Sect~\ref{sec:lowldef}) \\
$\hbox{\lowTEB} $& pixel-based temperature and polarisation low-$\ell$ \planck\ public likelihood (cf. Sect~\ref{sec:lowldef}) \\
$\hbox{\taureio} $& auxiliary constraint on \taureio\ from \planck\ reionisation measurement with \lollipop\ (cf. Sect~\ref{sec:lowldef}) \\
\hline
$\hbox{\VHL} $& very high-$\ell$ data (cf. Sect~\ref{VHL}) \\
$\hbox{\BAO} $& latest DR12 BAO data (cf. Sect~\ref{sec:baointro}) \\
$\hbox{\SN} $& JLA supernovae compilation (cf. Sect~\ref{sec:SN}) \\
\hline                  
\end{tabular}
\caption{Summary of data and likelihoods with their corresponding acronyms.  All are ready to use in the \CAMEL\ software. 
\Plik, \Commander, \lowTEB\ 
are available through the \planck\ PLA.}  
\label{tab:acro} 
\end{table*}

\subsection{\planck\ high-\texorpdfstring{$\ell$}{l} likelihoods}
\label{lik_intro}

In order to assess the impact of foreground residuals modelling on the \mnu\ constraints, we make use of different \planck\ high-$\ell$ likelihoods (\hlp\ and \plik).
They both use a Gaussian approximation of the likelihood based on cross-spectra between half-mission maps at the three lowest frequencies (100, 143 and 217\GHz) 
of \planck-HFI, but rely on different assumptions for modelling foreground residuals.
Comparing the results on \mnu\ obtained with both of 
these likelihoods is a way to assess a systematic uncertainty on the foreground residuals modelling.

\plik\ is the public \planck\ likelihood. It is described in detail in \citet{planck2014-a13}.
It uses empirically motivated power spectrum templates to model 
residual contamination of foregrounds (including dust, CIB, tSZ, kSZ, SZxCIB and point sources) in the cross-spectra.
The foreground residuals in \hlp\ are directly derived from \planck\ measurements \citep{Couchot:2016vaq}:
This is the main difference between \hlp\ and \plik.
For \lcdm\ cosmology, both likelihoods have been compared in~\citet{planck2014-a13}. 

In any of the \planck\  high-$\ell$ likelihoods, the residual amplitudes of the foregrounds 
are compatible with expectations, with only a mild tension on unresolved point-source amplitudes coming essentially from the 100 \GHz\ frequency \citep[we refer to Sect.~4.3 in][]{planck2014-a13}.
In order to assess the impact of the point-source modelling on the parameter reconstructions (and in particular \mnu), we use two variants of the \hlp\ likelihood. 
The first one, labelled \hlpPS, makes use of a physical model with two unresolved point-source components, corresponding to the radio and IR frequency domains, with fixed frequency scaling factors and number counts tuned on data \citep{Couchot:2016vaq}. 
The second one, labelled \hlpTT, uses one free amplitude for unresolved point-sources per cross-frequency leading to six free parameters \citep[as used in][]{couchot:2015}, in a similar way as what is done in \plik. This allows one to alleviate the tension on the point-source amplitudes. Both \hlpPS\ and \hlpTT\ lead to very similar results in the $\baselambdaCDM$ model, with a lower level of correlation between parameters for the former. Comparing results obtained with \hlpPS\ and \hlpTT\ is therefore useful for assessing their robustness with respect to the unresolved point-source tension. 

Both \hlp\ and also\plik\ include polarisation information using the $\hbox{EE}$ and $\hbox{TE}$ angular cross-power spectra.
Unless otherwise explicitly stated, only the temperature ($\hbox{TT}$) part is considered in the following.

Together with auxiliary constraints on nuisance parameters (such as the relative and absolute calibration) associated to each likelihood, we can also add
a Gaussian constraint to the SZ template amplitudes as suggested in \citet{planck2014-a13}. This constraint is based on a joint analysis of the \planck-2013 data with those from \ACT\ and \SPT~(see Sect.~\ref{sec::vhldata}) and reads:
\begin{equation}
\label{Aszcoo} A_{\mathrm SZ} = A_{\mathrm kSZ} + 1.6A_{\mathrm tSZ} = 9.5\pm 3~\mu K^2
,\end{equation}
when normalized at $\ell=3000$. The role of this additional constraint is also discussed in the following.

\subsection{Low-\texorpdfstring{$\ell$}{l}}
\label{sec:lowldef}
At low-$\ell$, two options are investigated to study
the impact of one choice or another on the \mnu\ limit determination:
\begin{itemize}
\item{} \lowTEB A pixel-based likelihood that relies on the \planck\ low-frequency instrument 70~GHz maps for polarisation and on a component-separated map using all Planck frequencies for temperature (\Commander).
\item{} A combination of a temperature-only likelihood, \Commander~\citep{planck2014-a13}, based on a component-separated map using all \planck\ frequencies,
and a 
Gaussian auxiliary constraint on the reionisation optical depth, $$\taureio=0.058\pm0.012\ , $$
derived from the last \planck\ results of the reionisation optical depth~\citep{planck2014-a25} \lollipop\ likelihood~\citep{lollipop}.
\end{itemize}

\subsection{High-resolution CMB data}
\label{sec::vhldata}
High resolution CMB data, namely the \ACT, \spthigh,
and \sptlow\ datasets are also used in this work. They  are later quoted ``\VHL'' (Very high-$\ell$) when combined altogether. 
The  \ACT\ data are those presented in
\cite{Das:2014}. They correspond to cross 
power spectra between the 148 and 220\GHz\ channels built from observations performed on two different sky areas
(an equatorial strip of about $300\ \mathrm{deg}^2$ and a southern strip of $292\ \mathrm{deg}^2$ for the 2008 season, and about $100\ \mathrm{deg}^2$ 
otherwise)   and during several seasons (between 2007 and 2010), for
multipoles between 1000 and 10000 (for 148$\times$148) and 1500 to 10000
otherwise.
For \SPT, two distinct datasets are examined. 
The higher $\ell$ part, dubbed \spthigh, implements the results, described in 
 \cite{Reichardt:2012}, from the observations of $800\ \mathrm{deg}^2$ 
at 95, 150, and 220\GHz\ of the SPT-SZ survey. The cross-spectra cover the $\ell$ range
between 2000 and 10000. As in~\cite{couchot:2015}, we prefer not to consider the more recent data from~\cite{George:2014} 
because the calibration, based on the \planck\ 2013 release, 
leads to a 1\% offset with respect to the last \planck\ data. 
We also add the \citet{Story:2012} dataset, dubbed \sptlow,
 consisting of a 150\GHz\ power spectrum, which ranges from $\ell=650$  to 3000, resulting from the analysis of 
observations of a field of $2540\ \mathrm{deg}^2$. 
Both SPT datasets have an overlap in terms of sky coverage and frequency. We have however checked that this
did not bias the results by, for example, removing the 150x150 GHz part from the \spthigh\ likelihood, as was 
done in~\cite{couchot:2015}.

\subsection{\planck\ CMB Lensing}
\label{lensing}

The full sky CMB temperature and polarisation distributions are
inhomogeneously affected by gravitational lensing due to large-scale structures. This is reflected in additional correlations
between large and small scales, and, in particular,  
in a smoothing of the power spectra in TT, TE, and EE.  
From the reconstruction of the four-point correlation functions~\citep{2002ApJ...574..566H},
one can reconstruct the
power spectrum of the lensing potential $C_\ell^{\phi\phi}$ 
of the lensing potential $\phi$.
In the following we make use of the corresponding 2015
temperature lensing  likelihood estimated by \planck~\citep{planck2014-a17}.

\subsection{Baryon acoustic oscillations}
\label{sec:baointro}
In Sect.~\ref{SNBAO}, information from the late-time evolution of the Universe geometry
are also included. 
The more accurate and robust 
constraints on this epoch come from the \BAO\ scale evolution. 
They bring cosmological parameter constraints that are highly complementary with those extracted from CMB, 
as their degeneracy directions are different.

\BAO\ generated by acoustic waves in the primordial fluid can be accurately 
estimated from the two-point correlation function of galaxy surveys. 
In this work, 
we use 
the acoustic-scale distance ratio $D_{\hbox{\tiny{V}}}(z)/r_{\mathrm{drag}}$
measurements 
from the 6dF Galaxy Survey at $z=0.1$~\citep{Beutler:2014yhv}.
At higher redshift, we included the 
BOSS DR12 BAO measurements that recently have been 
made available~\citep{Alam:2016hwk}. They consist in constraints on $ (D_M(z), H(z), f\sigma_8(z)) $ 
in three redshift bins, which encompass both BOSS-LowZ  and   BOSS-CMASS DR11 results. 
Thanks to the addition of the results on $f\sigma_8(z)$
the constraints on \mnu\ are significantly reduced with respect to 
previous  \BAO\ measurements~\citep{Alam:2016hwk}.
The combination of those measurements is  labelled  ``\BAO'' in the following.
We note that this is an update of the \BAO\ data with respect
those used in \cite{planck2014-a15}.

\subsection{Type Ia supernovae}
\label{sec:SN}
\SN\ also constitute a powerful
cosmological probe. The study of the evolution of their apparent magnitude with redshift 
played a major role in the discovery of late-time acceleration of the Universe.
We include the JLA compilation \citep{Betoule_2014}, which spans a
wide redshift range (from 0.01 to 1.2), while compiling up-to-date photometric data.
This is further referenced to as ``\SNplot'' in the following.

\section{CMB temperature results}
\label{cmbtemponly}
\subsection{Orders of magnitude}

\begin{figure}
\centering
\includegraphics[width=.49\textwidth]{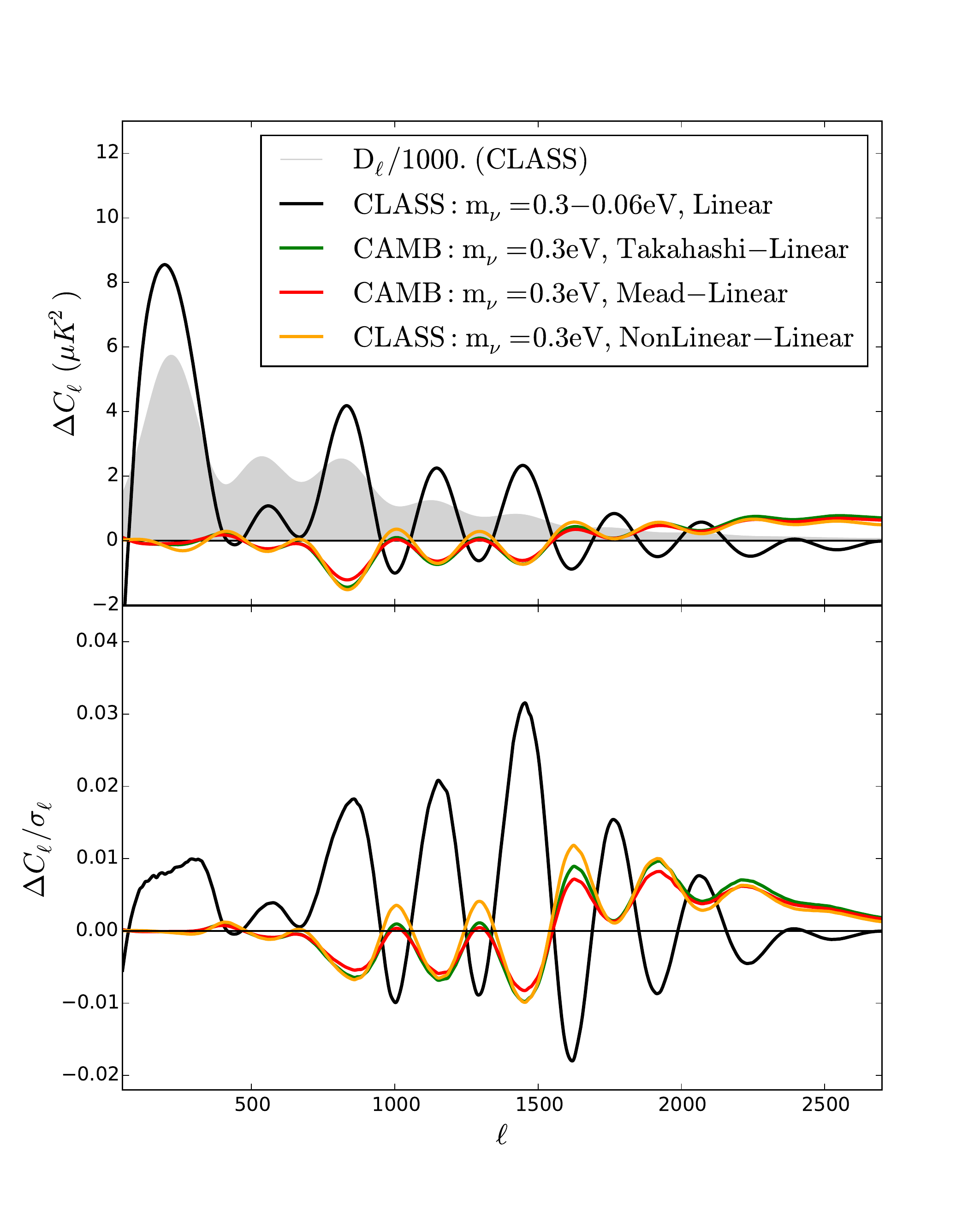}
\caption{\label{fig:magnitudeTT} Top: Absolute difference between the expected \troisnulambdaCDM\ TT  CMB spectrum
and a spectrum with the same values of the cosmological parameters except for \mnu=0.3\eV\ (computed with \CLASS\ (black)) in the
linear regime. The  shaded area is the original 
\troisnulambdaCDM\ spectrum rescaled to 1/1000. The differences introduced by the non-linear effects for \mnu=0.3\eV\ are shown for \CLASS\ in orange and 
\CAMB\ in red and green (cf. text). Bottom: The same differences relative to the uncertainties of the \hlpTT\ spectrum are shown. }
\end{figure}

The differences between the expected $C_\ell$ spectra for \mnu=0.3\eV\ and 
\mnu=0.06\eV\ in the \troisnulambdaCDM\ model are shown on Fig.~\ref{fig:magnitudeTT} 
in black on the upper panel without considering any non-linearities. 
The shaded area indicates the CMB spectrum divided by a factor $10^{3}$. 
The size of the effect of increasing \mnu\ up to 0.3\eV, except at the first peak,  is of the order of $\simeq 3\mu K^2$.
More interesting is the bottom part of this Figure (with the same color-code) 
where this difference is divided by the uncertainties estimated on the \hlpTT\ spectra. It shows that a sensitivity of 
few percent of a $\sigma$ over all the $\ell$ range has to be achieved in order to fit
for a 0.3\eV\ neutrino mass (the example taken here). 

The extreme case of the differences between linear and non-linear models of the CMB temperature power spectrum 
are also illustrated for \mnu=0.3\eV: For \CLASS, in orange, corresponding to~\cite{2012MNRAS.420.2551B},
and for \CAMB; where two models are compared, \mead\ in red
and \taka\  in green (cf. Sect.~\ref{designations}).
The plots show that the non-linear effects are of the order of 1$\mu$K and correspond to, at most, $\simeq 1\%$ of a $\sigma$.
The difference between those estimations gives a hint towards the theoretical uncertainty associated
to the propagation of non-linear effects. 
In addition to this, it must be kept in mind that when constraining extensions of $\lambdaCDM$ models, 
all the cosmological parameters are correlated, such
that those very small effects have to be disentangled from any other (more or less degenerated)
parameter's configuration.

To conclude, the effect one tries to fit on temperature power spectra
to extract information on \mnu\ is very tiny, and spreads over the whole multipole range. It therefore requires one to master  the 
underlying model used to build the CMB likelihood function to
a very high accuracy.

\subsection{\texorpdfstring{\MnutroisnulambdaCDM}{MnuLambdaCDM}}
\label{sec:mnlcdm}
\begin{figure}
        \begin{centering}
        \includegraphics[width=.49\textwidth]{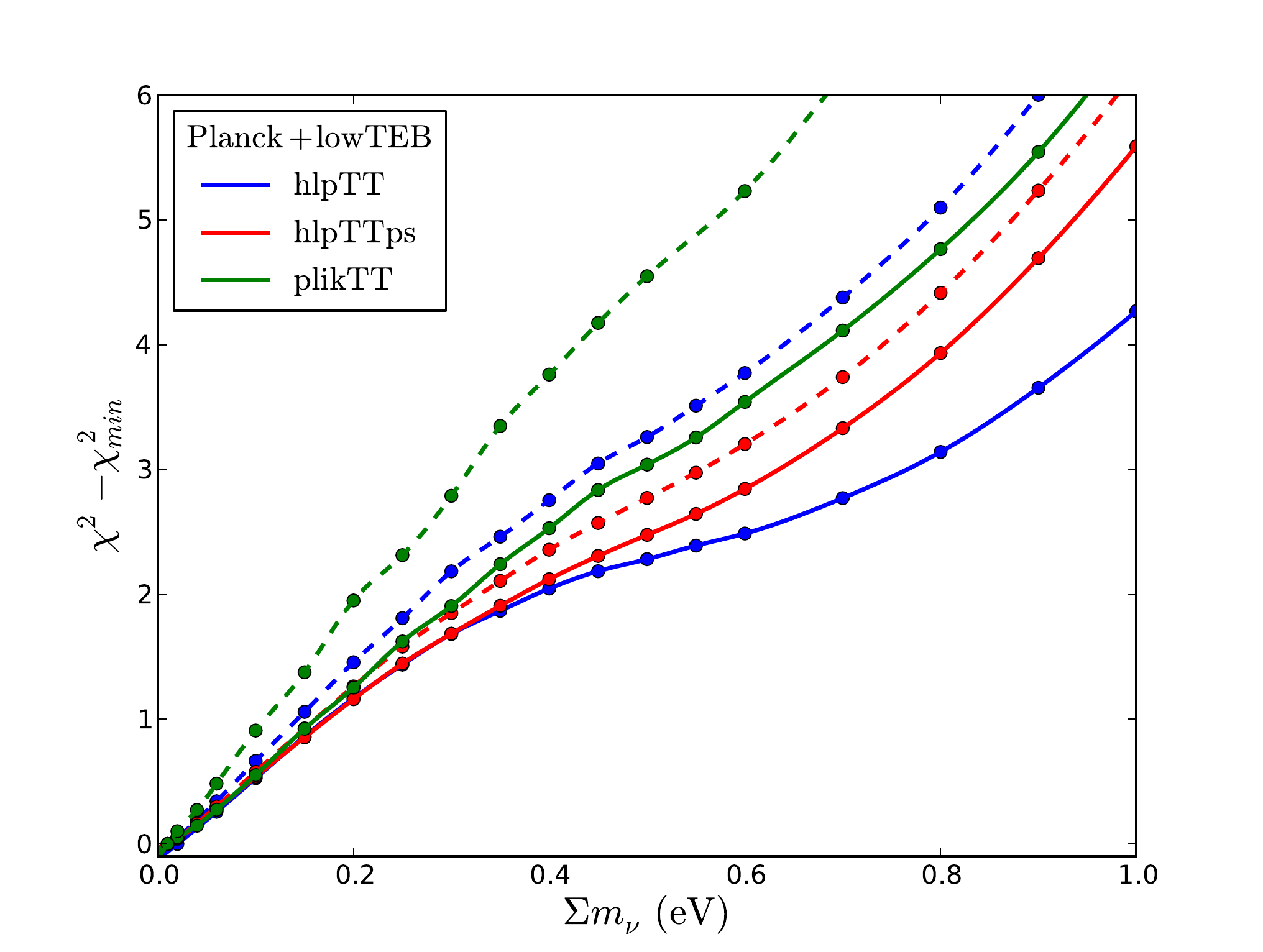}
        \caption{\mnu\ profile likelihoods obtained with \hlpTT\ (blue), \hlpPS\ (red), and \PlikTT\ (green) combined with \lowTEB (solid lines). The dashed lines include the constraint on the SZ amplitude (see Sect.~\ref{lik_intro}).}
        \label{fig:PlanckLowL}
        \end{centering}
\end{figure}

The profile likelihood results on \mnu\  derived from the 2013 \planck\ temperature power spectra have been
compared with those obtained with a Bayesian analysis in~\cite{planck2013-XVI} in the \MnubaselambdaCDM\ model. 
It was then shown that the profile likelihood shape was non-parabolic.
We recover the same results with the 2015
data in the \MnutroisnulambdaCDM\ model: This is illustrated for different high-$\ell$ likelihoods combined with \lowTEB\ on 
Fig.~\ref{fig:PlanckLowL}.

Fig. 3 illustrates that the  behaviour of the $\Delta\chi^2$ as
a function of \mnu\ is almost independent of the choice of the likelihood. Still,
the spread of the profile likelihoods gives an indication of the systematic uncertainties linked to this choice.
For such particular shapes of
the profile likelihood, one cannot simply use the Gaussian confidence level intervals
detailed in~\cite{FC}: One should rely
on extensive simulations to properly build the corresponding Neyman construction~\citep{Neyman333},
and apply the FC ordering principle;
this is beyond the scope of this work. We do not therefore quote any limit 
for non-parabolic profile likelihood. 

The use of the \Asz\  constraint (cf. Eq.~\ref{Aszcoo}) does improve the constraint on \mnu This is further
discussed in Sect.~\ref{VHL},
together with the impact of the combination of the 
\VHL\ data.

\subsection{Impact of low-\texorpdfstring{$\ell$}{l} likelihoods}
\label{sec:impact_lowell} 
\begin{figure}
        \begin{centering}
        \includegraphics[width=.49\textwidth]{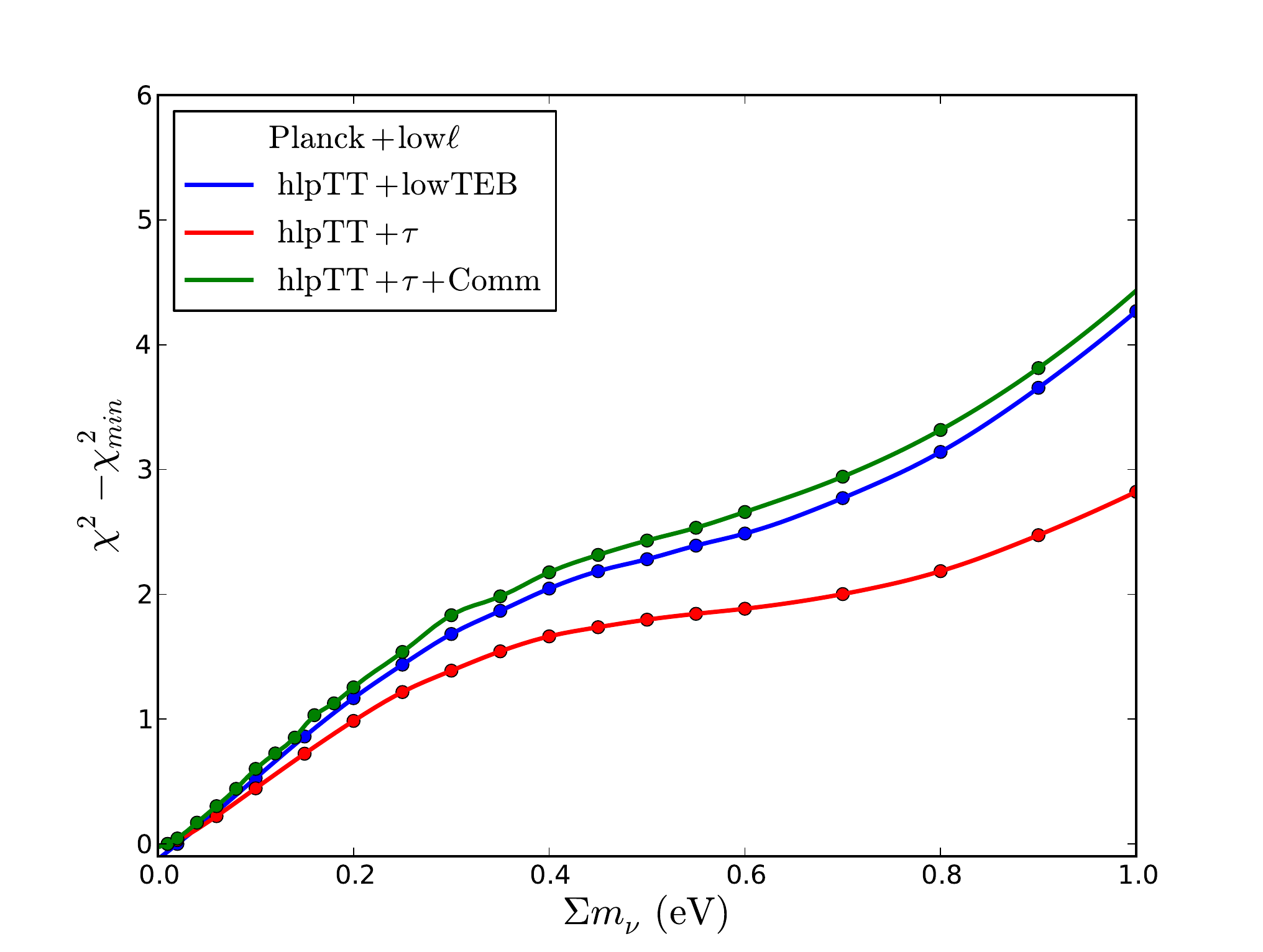}
        \caption{\mnu\ profile likelihoods obtained with \hlpTT, combined with different low-$\ell$ likelihoods: \lowTEB, and a \taureio\ auxiliary constraint combined or not with \Commander\
(see text for further explanation).}
        \label{fig:LowLNoVHL}
        \end{centering}
\end{figure}

%{\bf refaire le plot, mettre ce qu'on veut vraiment discuter et attention aux couleurs + section precedente}
On Fig.~\ref{fig:LowLNoVHL} are shown several \mnu\ profile likelihoods corresponding to different
choices for the low-$\ell$ likelihoods, while keeping \hlpTT\ for the high-$\ell$ part. They all present 
the same shape which, as previously, prevents us from extracting upper bounds.

The results obtained when combining \hlpTT\ with \lowTEB\ (in blue) are very close to those obtained 
with a \taureio\ auxiliary constraint+\Commander\ (in green), showing that with those datasets, the results do not significantly depend 
on the choice of the low-$\ell$ polarisation likelihood. 
%This could indicate that the main origin 
%of the particular shape of the profiles comes from the higher $\ell$ part.{\bf ??}
The same conclusion can
be derived from the comparison of the results obtained using  
\hlpTT+\taureio\ auxiliary constraint (in red). 

However, the difference between these two sets of profile likelihoods highlights the impact of 
 \Commander.  A possible origin of this difference lies in the fact that when adding \Commander 
in \troisnulambdaCDM+\alens, one reconstructs a  
higher \alens\ value. Indeed, with  \hlpTT+\taureio, we get $\alens=1.16\pm0.11$, while 
we find $\alens=1.20\pm 0.10$ for \hlpTT+\taureio+\Commander, that is, a higher value with a similar uncertainty. This higher 
tension with regards to the \lambdaCDM\ model (which assumes \alens=1) artificially leads to a tighter constraint on \mnu\ 
(we refer also to Sect.~\ref{alensAll}).

\subsection{Impact of \VHL\ data}
\label{VHL}
It was suggested in~\citet{planck2014-a13} to add a constraint on the SZ amplitudes
to mimic the impact of \VHL\ data, and we have shown in Fig.~\ref{fig:PlanckLowL}
that the use of such a constraint does tighten slightly 
the constraints on \mnu.

In this Section, 
we try to go one step further by actually using the \VHL\ data themselves 
to further constrain
the foreground residuals amplitudes in the \MnutroisnulambdaCDM\ case, using the
same procedure as the one described in~\cite{couchot:2015}.

\begin{figure}
\begin{centering}
\includegraphics[width=.49\textwidth]{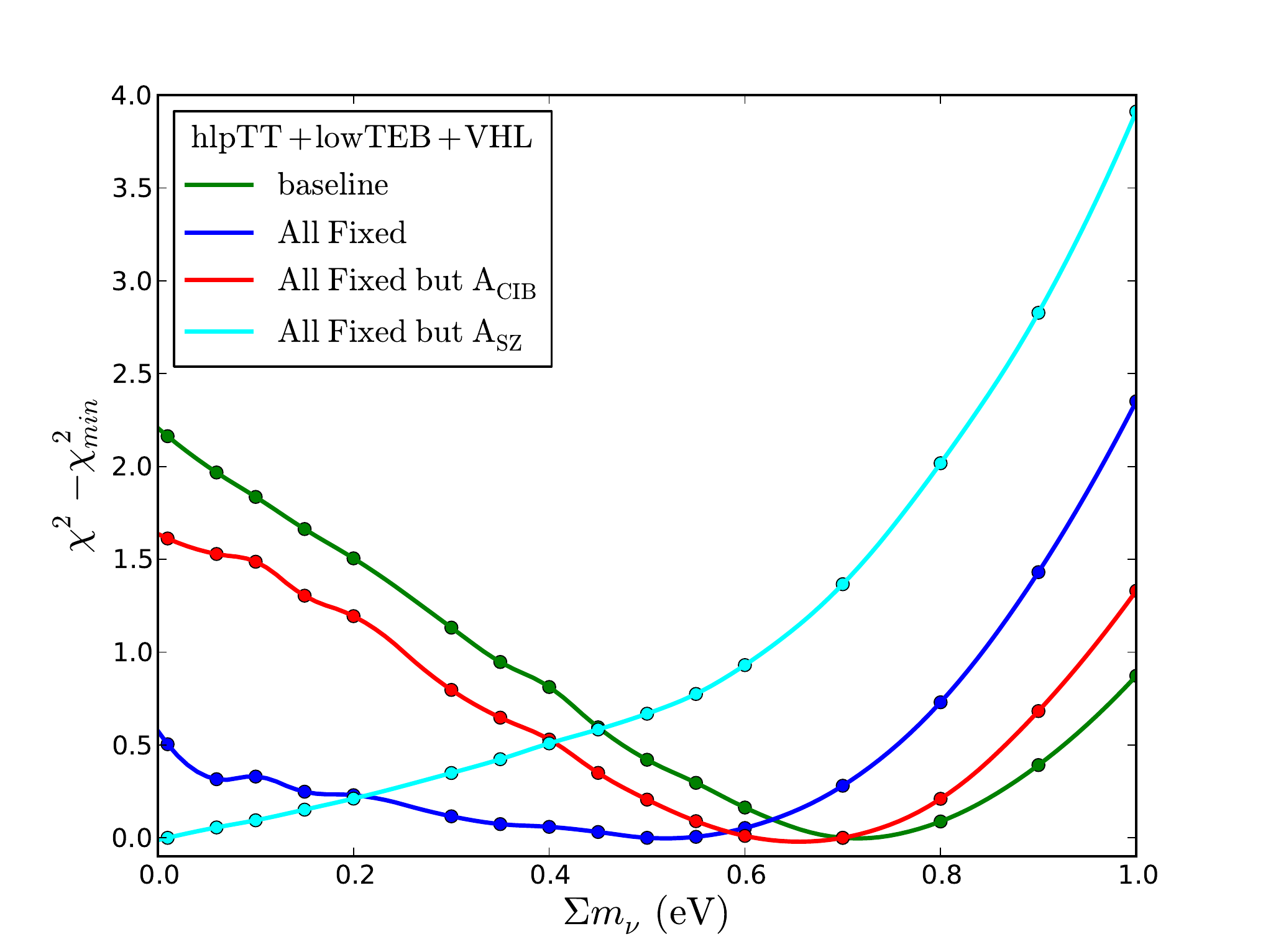}%prof_mnu_vhl_fg.pdf}
\caption{\mnu\ profile likelihoods  obtained for \hlpTT+\lowTEB+\VHL\ built with different  settings for the foregrounds: 
When fitting for all the foreground parameters, as usual, 
in green,  fixing the foreground nuisance parameters to their respective expected central values (and fixing $\Aksz$ and $\Aszcib$ to 0) in blue. 
We also report the profile likelihoods obtained when releasing one of the main foreground nuisance parameters at a time (cyan:~$\Asz$, red:~$\Acib$). 
\label{fig:profs_fg_effect}}
\end{centering}
\end{figure}

Fig.~\ref{fig:profs_fg_effect} shows the \mnu\ profile likelihoods obtained when combining  \hlpTT+\lowTEB\ with \VHL\
data in green: An apparent $\Delta\chi^2 $ minimum shows up, around $\mnu\sim 0.7,$~\eV\ with a $\Delta\chi^2 $ 
decrease with regards to $\mnu=0$ of the order of two units.
This is quite different from the \planck\ only \mnu\ profile likelihoods  previously studied, even
when the \Asz\ constraint has been added (cf. Sect.~\ref{sec:mnlcdm}). 
In the \MnubaselambdaCDM\ model, we have checked that the shape of the profile is about the same
but for the minimum, which is around \mnu=0.4~\eV, close to the results
obtained by~\citet{DiValentino:2013mt,hou:2014}.

To investigate  
this particular behaviour, we must stress
that, for the combination of \planck\ with \VHL\ data, one needs to 
compute the CMB power spectra up to $\ell\simeq\ 5000$. 
We therefore need 
to control the foreground residuals modelling, the datasets intercalibration uncertainties, and 
the uncertainties on non-linear effects models
over a very broad range of angular scales. 

To tackle the issue of the foreground modelling, 
several settings have  been studied. 
They are represented on Fig.~\ref{fig:profs_fg_effect}. 
The blue profile likelihood is  built
while fixing all the foreground amplitude nuisance parameters to their mean
expectation values. 
It can be compared with two other profile likelihoods (in cyan and in red), built 
when fitting  only the  SZ and the CIB templates amplitudes, respectively  
(these foregrounds dominate at the higher end
of the $\ell$ range). 
The observed variations, regarding both the $\chi^2$ rise at low \mnu\,  
and the \mnu\ value at the minimum, with respect to the default case (in blue), 
show that our combination of \planck\ and \VHL\ datasets is too sensitive 
to the foreground residuals modellings to be reliable for the derivation of a limit on \mnu. 
This may also come from the fact that the modelling
we have used for the full sky \planck\ surveys is not accurate enough
for the \VHL\ small patches of the sky.

We have also investigated the impact of the uncertainties on the modelisation of non-linear effects. 
The mean values of the cosmological parameters, derived from the best fits of the \hlpTT+\lowTEB+\VHL\
for \mnu=0.06~\eV\ and for 0.7~\eV, were used to compute the temperature $C_\ell$ spectra. 
We have observed that  the difference between these 
spectra was of the same order
of magnitude as the difference of spectra 
expected from two non-linear models for \mnu=0.06~\eV\ (namely between \taka\ and
\mead\ cf.~Sec.~\ref{sec:boltz}). As such a difference leads to a variation
of up to 2 $\chi^2$ units, we could expect that the uncertainties on non-linear models would lead to similar
$\chi^2$ differences\footnote{Still, the proper propagation of the uncertainties of non-linear
effects is beyond the scope of this work.}.
In addition, it must be noted that this difference is also of the order of magnitude
of the relative calibration between the different VHL datasets and \planck.

For all those  reasons, we have chosen not to include the 
\VHL\ datasets in the following (we refer also to~\cite{Addison:2015wyg} 
for the tensions between VHL datasets and \planck).
The potential impact of the uncertainties on non-linear models becomes
negligible when one only considers CMB spectra up to $\ell\simeq 2500$ (e.g. for \planck-only data).

\section{Adding \BAO\ and \SN\ data} 
\label{SNBAO}

\begin{figure}[!htbp]
        \centering
        \includegraphics[width=.49\textwidth]{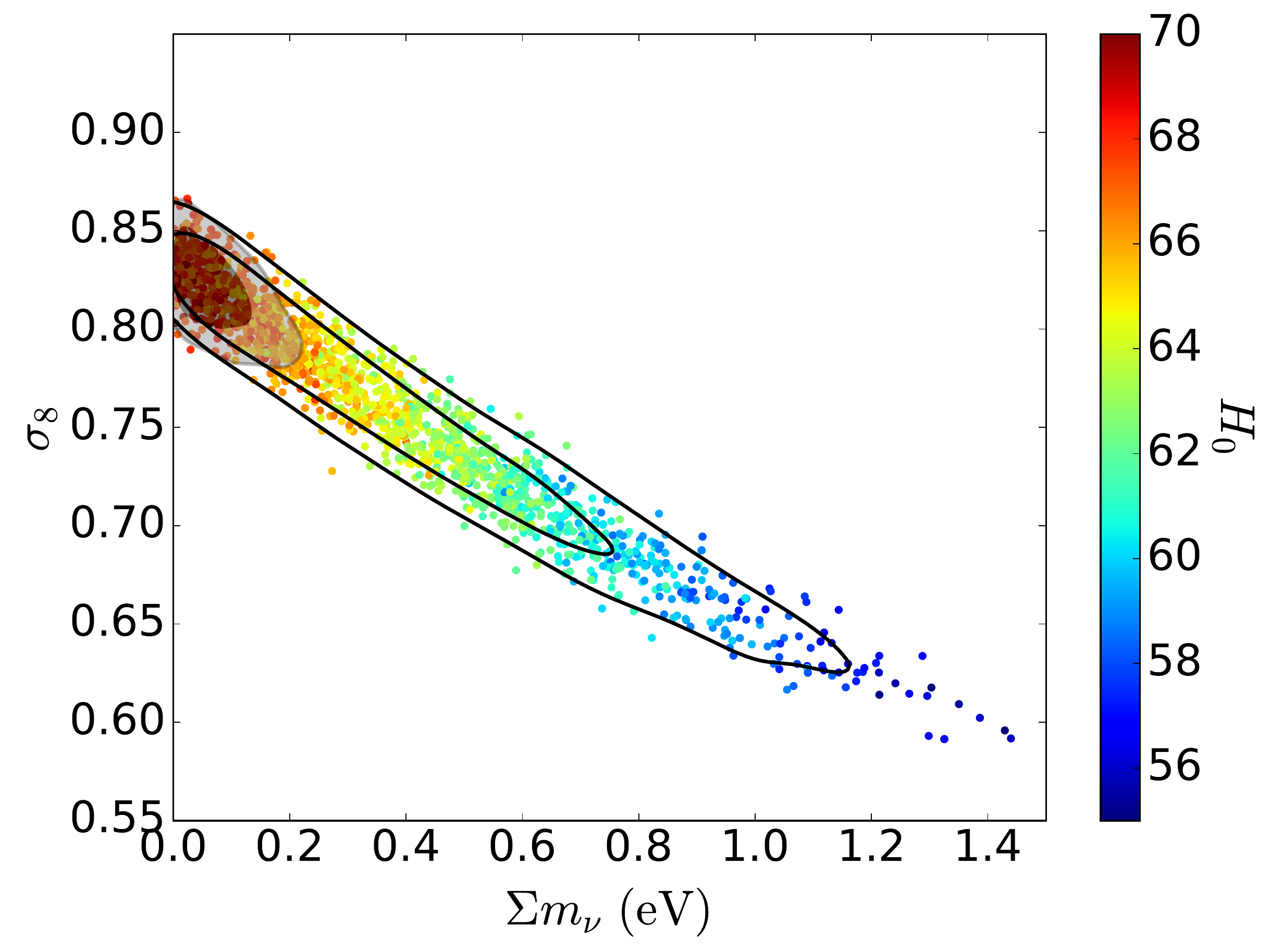}
        \caption{Bayesian sampling of the \hlpPS+lowTEB posterior in the $\mnu$--$\sigma_8$ plane, colour-coded by $H_0$. In flat \MnulambdaCDM\ models, higher \mnu\ damps $\sigma_8$, but also decreases $H_0$. Solid black contours show one and two $\sigma$ constraints from \hlpPS+lowTEB, while filled contours illustrate the results after adding BAO and SNIa data.}
        \label{fig:mnu_H0_sig8}
\end{figure}

As noted in Sect.~\ref{sec:nu_degeneracies}, the main degeneracy when using CMB data to constrain flat \MnulambdaCDM\ models, is between $\mnu$ and $\Omega_\Lambda$ which  are both related to the angular-diameter distance to the last scattering surface. This translates into a degeneracy between \mnu\ and the derived parameters $\sigma_8$ and $H_0$ as illustrated in Fig.~\ref{fig:mnu_H0_sig8}.
The effect of neutrino free-streaming on structure formation favours lower $\sigma_8$ values at large \mnu, which in addition require one to lower $H_0$.
Adding BAO and SNIa data breaks this relation, and substantially tightens the constraint on \mnu. 
%This permits
%to check the overall consistency of cosmological data, including a comparison with direct measurements of $H_0$. 
In this section, we analyse the combination of \planck\ CMB data with DR12 \BAO\ and \SN\ data (as described in Sect.~\ref{designations}).

%As noted in Sect.~\ref{sec:baointro}, combining late probes and CMB measurements
%helps breaking degeneracies between parameters, therefore improving the accuracy on
%the cosmological parameters constraints.  
%In the following, we analyze the combination of  \planck\ temperature 
%with DR12 \BAO\ and \SN\ data (cf. Sect.~\ref{designations}). The
%results presented in this Section are therefore using an update of
%\BAO\ data if one wants to compare with~\cite{planck2014-a15}.

\subsection{\hlpTT, \hlpPS,\ and \PlikTT\  comparison}
\label{sec:mnu_interm_ell}

\begin{figure}
        \centering
        \includegraphics[width=.49\textwidth]{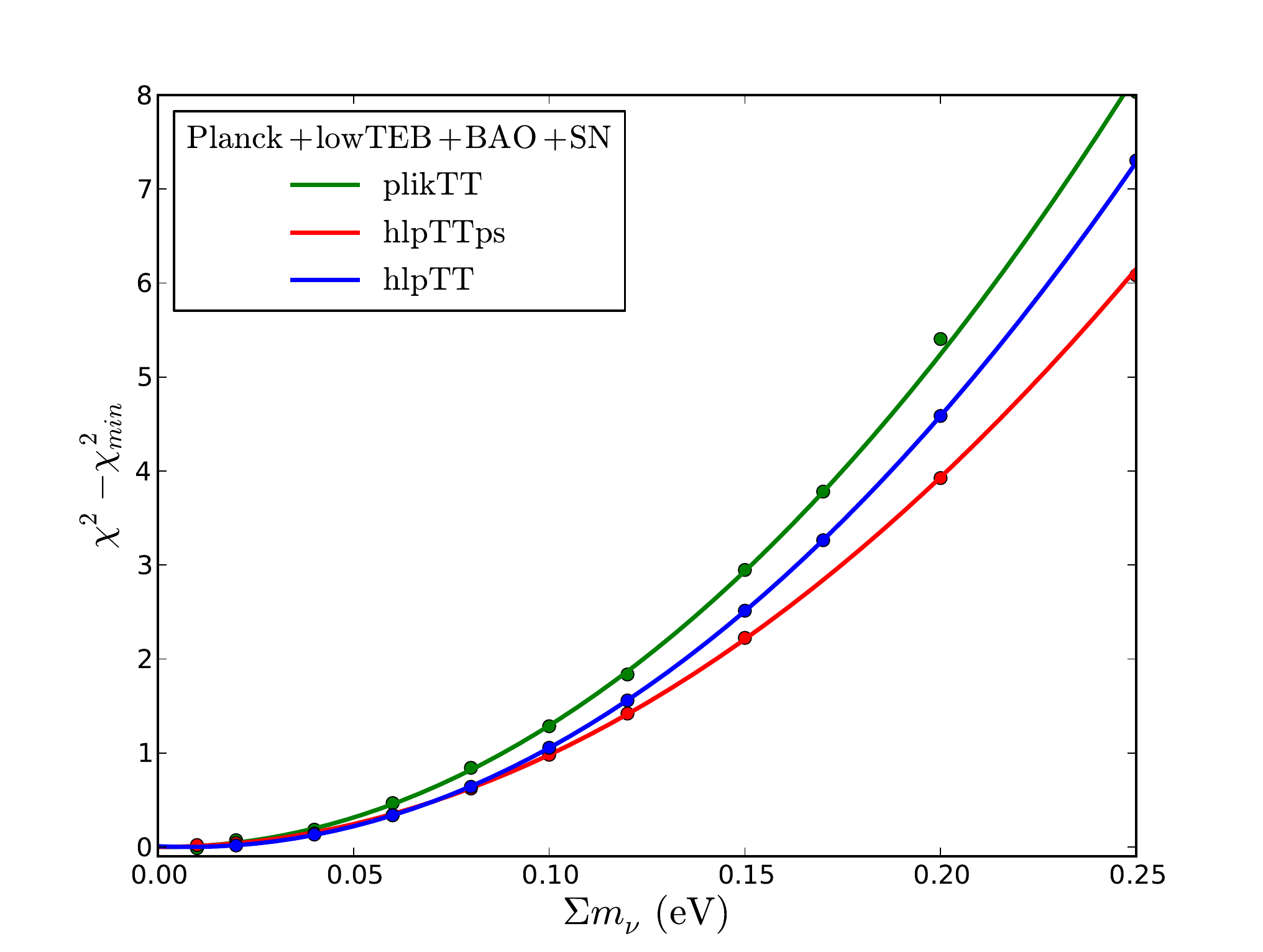}
        \caption{\mnu\ profile likelihoods derived for the combination of \lowTEB, various \planck\ high-$\ell$ likelihoods, \BAO\ and \SN: A comparison is made between \hlpTT, \hlpPS, and \PlikTT.} 
         \label{fig:baosn}
\end{figure}

Fig.~\ref{fig:baosn} compares the three \planck\ likelihoods when they are
combined with \lowTEB, \BAO\ and \SN. The impressive improvement with respect 
to the \planck\ only results (Fig.~\ref{fig:PlanckLowL}) can be measured, for example, by the comparison
of the range of \mnu\ values for which the $\Delta\chi^2$ is below 3.
As expected, those results illustrate
that most of the constraint on \mnu\ does not come 
from CMB-only data (at decoupling neutrinos act essentially as radiation) 
but from the combination with late-time probes (where they contribute as matter).
In addition, for this combination of probes, 
the likelihood profiles take on a standard parabolic shape:
The derived upper bounds on \mnu, using the F.C. prescription,
are summarised in  Table~\ref{tab:limits_baosn}. 
We also quote the \alens\ values obtained using the same datasets for the \lambdaCDM+\alens\ model (fixing $\mnu=0.06 \eV$). We note that they differ from one by roughly 2$\sigma$. The impact on the \mnu\ limit is discussed in Sect.~\ref{alensAll}.

The profiles of the different high-$\ell$ likelihoods are very similar, giving confidence in the final results
that can be derived from their comparison. 
The spread between the curves reflects the remaining systematic uncertainty linked
to the choice of the underlying foreground modelling.
We have checked that, for \hlpTT\ and \hlpPS, removing the foreground nuisance
parameter auxiliary constraints does not impact the results: 
This provides an additional proof 
that the model and the data are in very good agreement.
The information added by the \Asz\ constraint is of no use in this particular combination
of data within the \MnutroisnulambdaCDM\ model.
The systematic uncertainty on the \mnu\ limit due
the foreground modelling, deduced from this comparison, is therefore estimated to be of the order 
of 0.03~\eV\ for this particular data combination.

As expected, the main improvement with respect to the \planck\ only case comes from
the addition of the \BAO\ dataset: the contribution on the \mnu\ limit of the addition of \SN\ 
is of the order of $\simeq 0.01~\eV$.

%WithBaoSN_3massive.py

\begin{table}
\centering          
\begin{tabular}{l c c } 
\hline\hline       
\hbox{\planck TT+\lowTEB} &  \mnu   & \alens \\
\hbox{\BAO+\SNplot} & limit ($\hbox{eV}$) & \\
\hline                    
$\hbox{\hlpTT }$& 0.18 & 1.16$\pm$0.09\\ 
$\hbox{\hlpPS}$& 0.20  & 1.14$\pm$0.08\\ 
$\hbox{\PlikTT} $& 0.17& 1.19$\pm$0.09 \\ 
%$\hbox{\PlikTT\ SZ\ prior} $&  0.17 &  -- \\ 
\hline                  
\end{tabular}
\caption{95$\%$ CL upper limits on \mnu\ in  \MnutroisnulambdaCDM\ (i.e. with $\alens=1$) and
results on \alens\  (68$\%$ CL) in the \troisnulambdaCDM+\alens\ model (i.e. with $\mnu=0.06 \eV$)
obtained when combining the \planckTT+\BAO+\SNplot.}  
\label{tab:limits_baosn} 
\end{table}

\subsection{Impact of low\texorpdfstring{-$\ell$}{l} likelihoods}
\label{sec:mnu_low_ell}
While in the previous Section  we focused on the estimation of the remaining systematic uncertainties linked
to the choice of the high-$\ell$ likelihood, a comparison of the low-$\ell$ parts is now performed.
We already discussed in Sect.~\ref{sec:impact_lowell} the impact 
of this choice on the results derived from CMB data only;
this comparison focuses on the combination of \BAO\ and \SN\ data.  

\begin{figure}
        \begin{centering}
        \includegraphics[width=.49\textwidth]{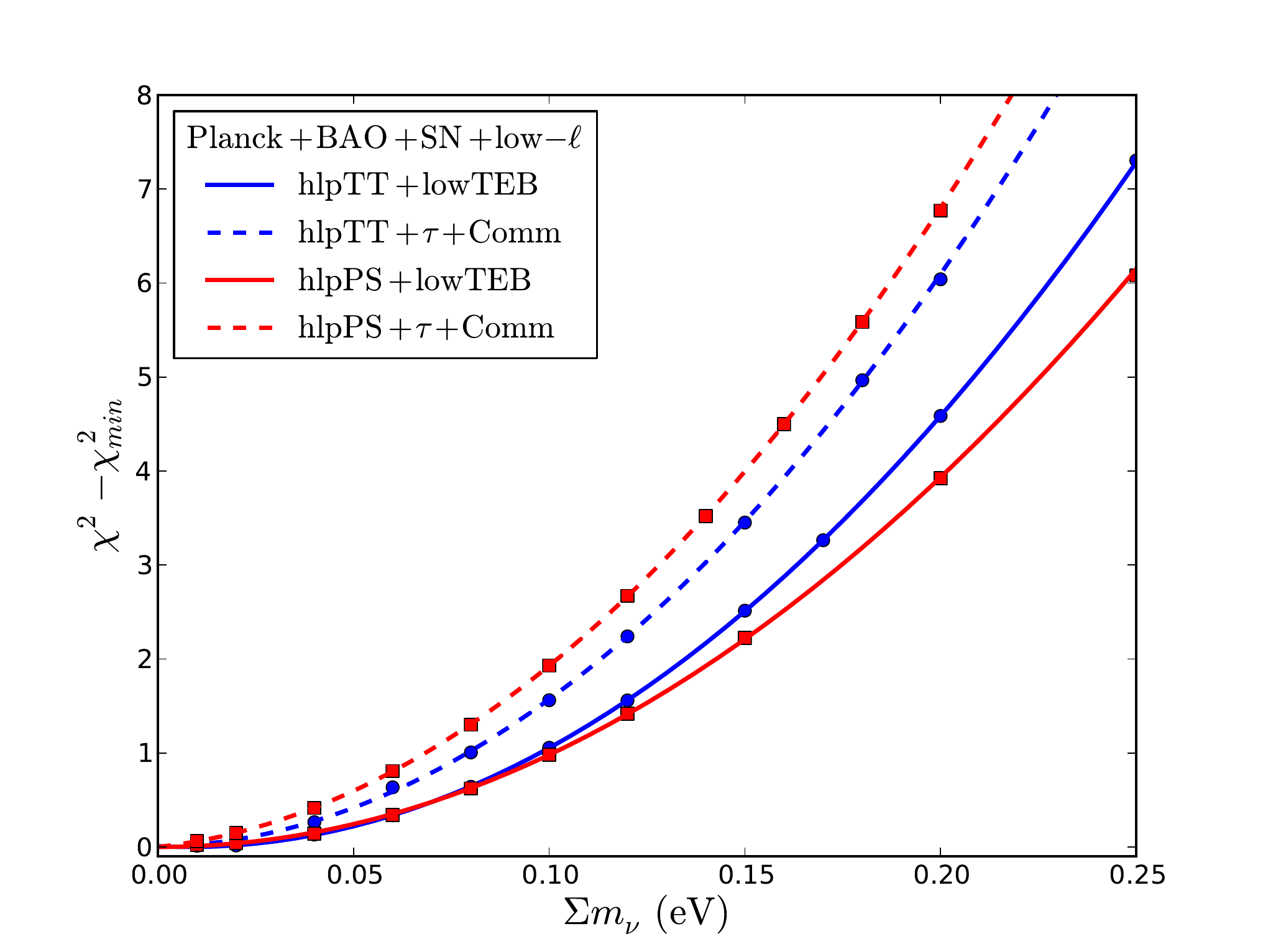}
        \caption{\mnu\ profile likelihoods derived for the combination of \planck\ high-$\ell$ likelihoods (\hlpTT\ and \hlpPS) with \BAO\ and \SN, and either \lowTEB\ or the $\tau$ auxiliary constraint at low-$\ell.$}
        \label{fig:all_lowell}
        \end{centering}
\end{figure}

The results are summarised in Fig.~\ref{fig:all_lowell}. 
For the two  \hlp\ likelihoods, tightening the constraints on \taureio\ 
with the use of 
\taureio+\Commander\  in place of \lowTEB  results in a limit of
 0.15\eV\ (resp. 0.16\eV) for \hlpPS\ (resp. \hlpTT)
and amounts to a few $10^{-2}$\eV\ decrease compared to the \lowTEB\ case. 
This decrease is a direct consequence of both the (\mnu,\taureio) correlation \citep{Allison:2015qca},
and the smaller value of the reionisation optical depth constraint from $\sim$ 0.07 to 0.058 \citep{planck2014-a25}. 

%cmp_lowells_3massive.py

\subsection{Cross-check with TE}
\label{sec:mnu_TE}
\begin{figure}
\centering
\includegraphics[width=.49\textwidth]{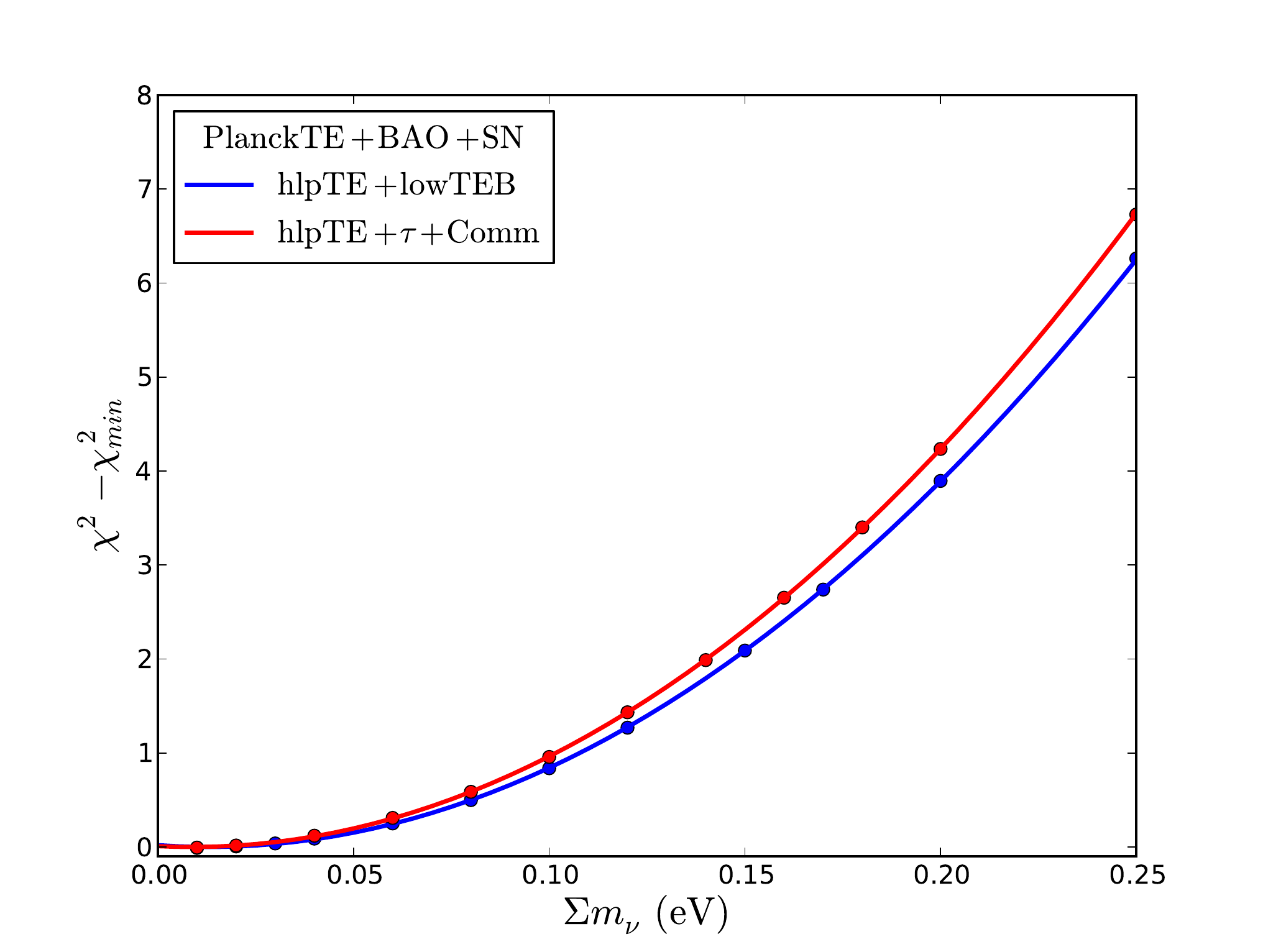}
\caption{\label{fig:mnu_TE} \mnu\ profile likelihoods obtained when combining \hlpTE\ with either \lowTEB\ (red), 
or an auxiliary constraint on $\taureio$+\Commander\ (blue) and with \BAO\ and \SNplot.}
\end{figure}

As pointed out in~\cite{galli:2014} and~\cite{Couchot:2016vaq},
CMB temperature-polarisation cross-correlations (TE) give
competitive constraints on \lambdaCDM\ parameters.  The leading advantage of using only these data is that one
depends very weakly on foreground residuals and therefore uncertainty linked to the model parametrisation is reduced. In practice, 
only one foreground nuisance parameter is required: The amplitude of the
polarized dust.
Nevertheless, the S/N being lower than in
the TT case for \planck, a likelihood based on TE spectra is not competitive when constraining
extensions to the six \lambdaCDM\ parameters.
Indeed an estimation of the TE-only constraint on \mnu\ would lead to a limit higher
than 1\eV. However, as soon as \BAO\ data are added, one obtains a
constraint competitive with TT as shown in Fig.~\ref{fig:mnu_TE}. As in the TT case, 
all profile likelihoods are nicely parabolic, and the corresponding
limits are summarised in Table~\ref{tab:limits_TE}.

\begin{table}
\centering          
\begin{tabular}{l c  } 
\hline\hline       
\hbox{\planck TE+low-$\ell$} &  \mnu    \\
\hbox{+\BAO+\SNplot} & limit ($\hbox{eV}$)  \\
\hline                    
$\hbox{\hlpTE+\lowTEB }$& 0.20\\ 
$\hbox{\hlpTE+}\taureio+$\Commander& 0.19\\ 
\hline                  
\end{tabular}
\caption{95$\%$ CL upper limits on \mnu\ in  \MnutroisnulambdaCDM\ 
obtained with \hlpTE+\BAO+\SNplot\ in combination with
\lowTEB, or an auxiliary constraint on $\taureio$ and \Commander.}  
\label{tab:limits_TE} 
\end{table}

As for temperature-only data,
adding the \SNplot\ data improves only very marginally the results 
up to 0.01~\eV.
Tests of the dependencies on the low-$\ell$ likelihoods have also been
performed and an example is given in Table~\ref{tab:limits_TE}.
As a final result, we obtain \mnu<0.20~\eV\  at 95\% CL as strong as in the TT case,
showing that the loss in signal over noise of TE (statistical uncertainty) 
is balanced by improved control of foreground modelling (systematic uncertainty).

\subsection{\texorpdfstring{\alens}{Alens}\ and \texorpdfstring{\mnu}{Mnu}}
\label{alensAll}

\subsubsection{\texorpdfstring{\MnutroisnulambdaCDM\ model}{MnutroisnulambdaCDM}}

As previously stated, CMB data tend to favour a 
value of \alens\ greater than one.
In the combination of \planck\ high-$\ell$ likelihood with \lowTEB, \BAO\ and
\SN, the \alens\ values estimated in the \troisnulambdaCDM+\alens\ model, 
are summarised in the third column of Table~\ref{tab:limits_baosn}.
As expected they are almost identical to the ones obtained with CMB
data only. 

The fact that \alens\ is 
not fully compatible with the \lambdaCDM\ model, 
has to be taken into account when
stating final statements on \mnu\ since, otherwise, the 
results are not obtained within a coherent model: On one side
we fix \alens\ to one by working within a \MnulambdaCDM\ model while the
data are, at least, $\simeq2\sigma$  away from this value, and on the other side,
fixing $\alens=1$ results, artificially, in a tighter constraint on \mnu. 
This last effect can be seen, for example, in Table~\ref{tab:limits_baosn},
for which the higher the \alens\ value, the tighter the constraint on \mnu.

There are two ways to propagate this effect on the \mnu\ limit determination. 
The first is to open up the parameter space to \MnutroisnulambdaCDM+\alens\ (as it is done in the Sect.~\ref{sec:mnu+alens}).
The second is to better constrain the lensing sector by considering the \planck\ lensing likelihood and then to fit only for the \mnu\ extension using the \MnutroisnulambdaCDM\ model, fixing $\alens=1$ (cf. Sect.~\ref{sec:lens1}).

\subsubsection{\texorpdfstring{The \MnutroisnulambdaCDM+\alens\ model}{MnutroisnulambdaCDM and alens}}
\label{sec:mnu+alens}
In this Section, we open the \MnutroisnulambdaCDM\ parameter space to \alens\ 
for the combination of \planck\ high-$\ell$ likelihoods
with lowTEB+\BAO+\SN.

The limits derived from the corresponding profile likelihoods are summarised in Table~\ref{tab:limits_alens}.
The increase of the limits with respect to those of Table~\ref{tab:limits_baosn} results from two 
effects. First of all we open up the parameter space, propagating the uncertainty on 
\alens\ on the \mnu\ determination. 
The second effect is linked to the fact that, as already stated, 
the CMB data tend to favour a higher \alens\ value than
expected within a \lambdaCDM\ model. We have observed 
that 
this effect propagates as an increase of the baryon energy density, a slight decrease of the cold dark matter energy density,
and this shows up, with a fixed
geometry, as a higher neutrino energy density. Those two combined effects
drive
the limit to high values of \mnu\ when fitting for both \mnu\ and \alens.\  

\begin{table}[!h]
\centering          
\begin{tabular}{l c } 
\hline\hline       
\hbox{\planck TT+\lowTEB} &  (\mnu [eV],\Alens) \\
\hbox{\BAO+\SNplot} &   \\
\hline                    
$\hbox{\hlpTT}$&  (0.39,~$1.22 \pm 0.12$)   \\
$\hbox{\hlpPS}$&  (0.34,~$1.18 \pm 0.10$) \\ 
$\hbox{\PlikTT} $&  (0.40,~$1.28 \pm 0.12$)   \\ 
\hline                  
\end{tabular}
\caption{Results on \mnu\ (95$\%$ CL upper limits) and \alens\ (68\% CL) obtained from a combined fit in the \MnutroisnulambdaCDM+\alens\ model with \planckTT+\BAO+\SN.}
\label{tab:limits_alens}
\end{table}

\subsubsection{Combining with CMB lensing}
\label{sec:lens1}
Another way of tackling the \alens\ problem is to add the lensing 
\planck\ likelihood to the combination (see Sect.~\ref{lensing}).
This allows us to obtain a lower \alens\ value, as 
shown in the third column of Table~\ref{tab:limits_baosnTT_lens} 
in the \troisnulambdaCDM+\alens\ model.
With this combination, the \alens\ value extracted from the data is fully compatible with
the \lambdaCDM\ model, allowing us to derive a limit on \mnu\ together with a coherent \alens\ value.

As expected, in the \troisnulambdaCDM\ model, the \mnu\ limits are therefore pushed toward
higher values than what has been presented
in Table~\ref{tab:limits_baosn}: This is 
exemplified by the second column of Table~\ref{tab:limits_baosnTT_lens}.

\begin{table}[!h]
\centering          
\begin{tabular}{l c c } 
\hline\hline       
\hbox{\planck TT+\lowTEB} &  \mnu   & \alens \\
\hbox{\BAO+\SNplot+lensing} & \hbox{ limits (eV)} &  \\
\hline                    
$\hbox{\hlpTT }$        & 0.21  & 1.06 $\pm$ 0.05\\
$\hbox{\hlpPS}$ & 0.21  & 1.06 $\pm$ 0.06\\
$\hbox{\PlikTT}$        & 0.23  & 1.05 $\pm$ 0.06\\
\hline                  
\end{tabular}
\caption{95$\%$ CL upper limits on \mnu\ in  \MnutroisnulambdaCDM\ (i.e. with $\alens=1$) and
results on \alens\  (68$\%$ CL) in the \troisnulambdaCDM+\alens\ model (i.e. with $\mnu=0.06 \eV$)
obtained when combining \planckTT+\BAO+\SNplot+lensing.}  
\label{tab:limits_baosnTT_lens}
\end{table}

\subsection{Constraint on the neutrino mass hierarchy}
\label{sec:mnu_nhorih}

As explained in Sect.~\ref{sec:nu_degeneracies}, the neutrino mass repartition leaves a very small signature 
on the CMB and matter power spectra.  In this section, we test whether or not the combination of modern cosmological data is sensitive
to it.

We compare the results obtained with four configurations of neutrino mass settings.
The first one corresponds to one massive and two massless neutrinos as in \MnubaselambdaCDM\ and is
labelled $[1\nu]$. The second one is built under the assumption of three mass-degenerate 
neutrinos as in \MnutroisnulambdaCDM\ and is denoted $[3\nu]$.
We also discuss the normal hierarchy [3$\nu$ NH] (resp. inverted hierarchy [3$\nu$ IH]) 
derived from Eq.~\ref{eq:dm12} and Eq.~\ref{eq:dm13}  (resp. Eq.~\ref{eq:dm13ih}).

\begin{figure}[!ht]
        \centering
        \includegraphics[width=.49\textwidth]{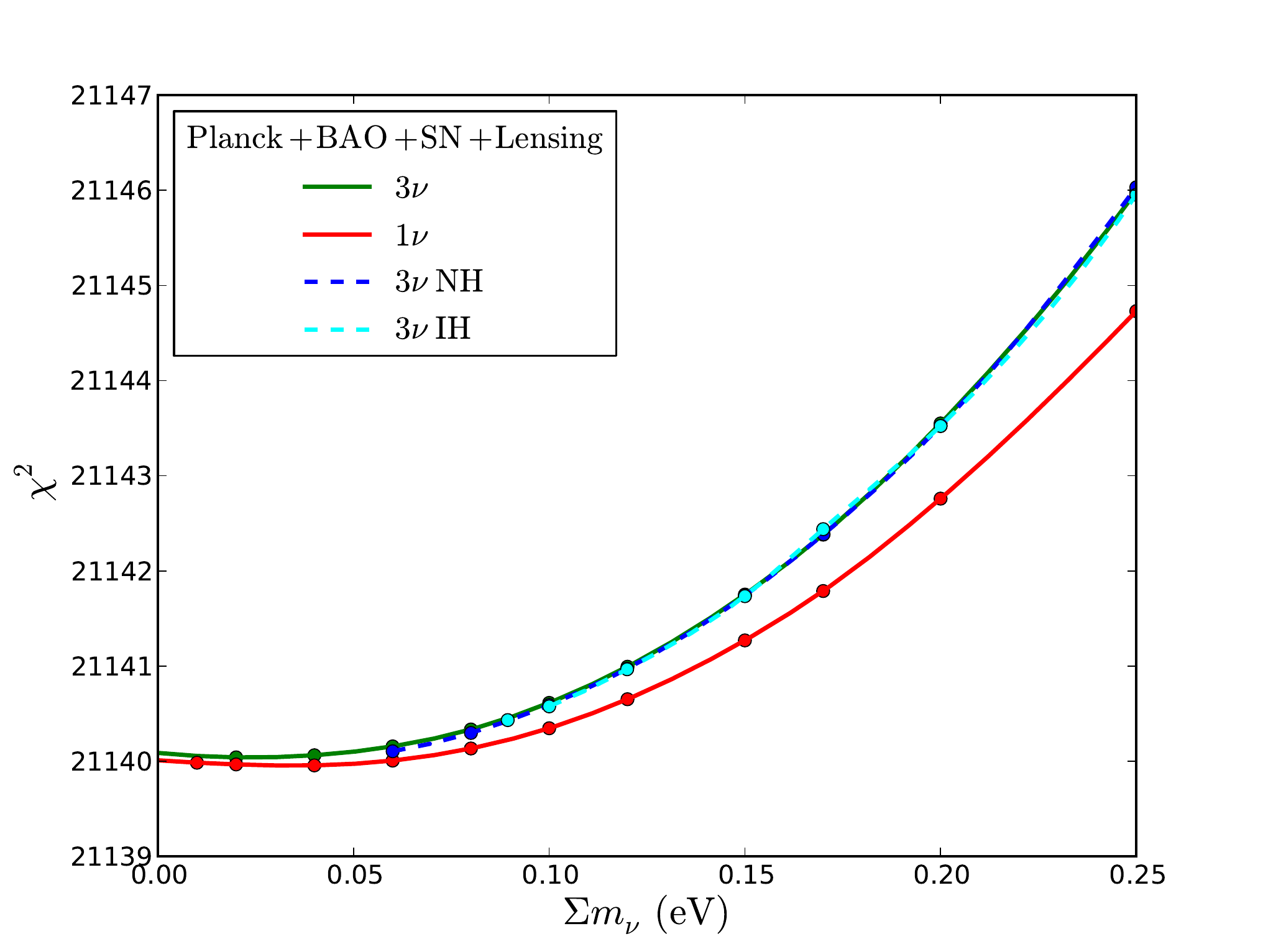}
        \caption{\label{fig:prof_nhih_lens} Profiled $\chi^2$ on \mnu\ derived for the combination \hlpTT+\lowTEB+\BAO+\SN+lensing in the one massive, two massless scenario (red), in the degenerate masses hypothesis (green), and for normal (NH, dashed blue line) 
and inverse (IH, dashed cyan line) hierarchies.}
\end{figure}

In contrast with the rest of this paper, we did not subtract, in this Section, the minimum of the $\chi^2$ 
to plot the
profile likelihoods. This allows us to assess the $\chi^2$ difference
between the various neutrino configurations. 
In Fig.~\ref{fig:prof_nhih_lens}, we show the results obtained using the combination \hlpTT+\lowTEB+\BAO+\SNplot+lensing.
The 95\% CL upper limits derived from these profile likelihoods are reported in Table~\ref{tab:limits_nhih}.

The observed difference between [1$\nu$] and [$3\nu$] illustrates the impact of the choice
of the number of massive neutrinos on the derived constraint on \mnu. 
More important is the comparison of the profile likelihoods built for the different hierarchy scenarios.
The fact that they are indistinguishable (both in shape and in absolute $\chi^2$ values),
and, even more, that they are almost identical
to the one of the three degenerated masses,
shows that there is, with modern data, no hint of a preference for the data towards one scenario or another, for this particular data 
combination~\citep[we
refer also to the latest discussion in][]{Schwetz:2017fey}.

\begin{table}[!h]
\centering          
\begin{tabular}{l c c } 
\hline\hline       
\hbox{$\nu$ mass setting} &  \mnu\ limits (eV) \cr
\hline
$[3\nu]\ \troisnulambdaCDM $& 0.21 \cr 
$[3\nu\ \hbox{NH}]$& 0.21 \cr 
$[3\nu\ \hbox{IH}]$ & 0.21 \cr 
$[1\nu]\ \baselambdaCDM$ & 0.23 \cr
\hline                  
\end{tabular}
\caption{95$\%$ CL upper limits on \mnu\ obtained with \hlpTT+\lowTEB+\BAO+\SN+lensing for different neutrino mass repartition: three degenerate masses, normal hierarchy (NH), inverse hierarchy (IH) and one massive plus two massless neutrinos.}
\label{tab:limits_nhih}
\end{table}

\section{Adding CMB polarisation}
In the previous Section, we derived limits on \mnu\  from various
high-$\ell$ \planck\ temperature likelihoods combined with \BAO\ and \SN. 
All those 
results were cross-checked with the almost foreground-free TE \planck\ spectra. 
In this Section, we combine the temperature and polarisation CMB data from \planck\
together with \BAO, \SN. As done previously, the 
CMB lensing is then also added in the combination to address the \alens\ tension. 
We then show the final results of this paper on the \mnu\ determination.

\subsection{Combination of TT, TE, EE \BAO\ and \SN}
\label{sec:mnu_All}

The 95\%\ CL upper limits on \mnu\  
corresponding to the full TT+TE+EE likelihoods (labelled ALL), combined with \BAO, \SNplot\ and \lowTEB\
are summarised in Table~\ref{tab:limits_baosnALL}.

They are very close to the temperature-only upper limit of Table~\ref{tab:limits_baosn}, 
showing that the use of the polarisation information in addition to the temperature does not add much information.
They are also very close, showing the consistency of the
results with respect to the high-$\ell$ \planck\ likelihoods when \BAO\ and \SN\ are included.

\begin{table}
\centering          
\begin{tabular}{l c } 
\hline\hline       
\hbox{\planck ALL+\lowTEB} &  \mnu \\
\hbox{+\SNplot+\BAO} & \hbox{ limits (eV)}   \\
\hline                    
$\hbox{\hlpALL }$& 0.18  \\
$\hbox{\hlpALLps}$& 0.18 \\
$\hbox{\plikALL} $& 0.15  \\
\hline                  
\end{tabular}
\caption{95$\%$ CL upper limits on \mnu\ in  \MnutroisnulambdaCDM\ 
obtained when combining \plikALL, \hlpALL\ or \hlpALLps\  with \lowTEB+\SNplot+\BAO. 
ALL refers to the combination TT+TE+EE. }
\label{tab:limits_baosnALL} 
\end{table}

However, for this data combination, we are still left with a 2$\sigma$ tension on \alens\
(the \alens\ values are almost the same as the ones of the TT combination of
Table~\ref{tab:limits_baosn}).
The fact that the results from \plikALL\ are lower than those of \hlp\
is linked to the fact that the \alens\ value of \Plik\ is higher than the one
of \hlp. We will come back to this point in the next Section.

\subsection{Combining with CMB lensing}
\label{lens2}
\begin{figure}
\centering
\includegraphics[width=.49\textwidth]{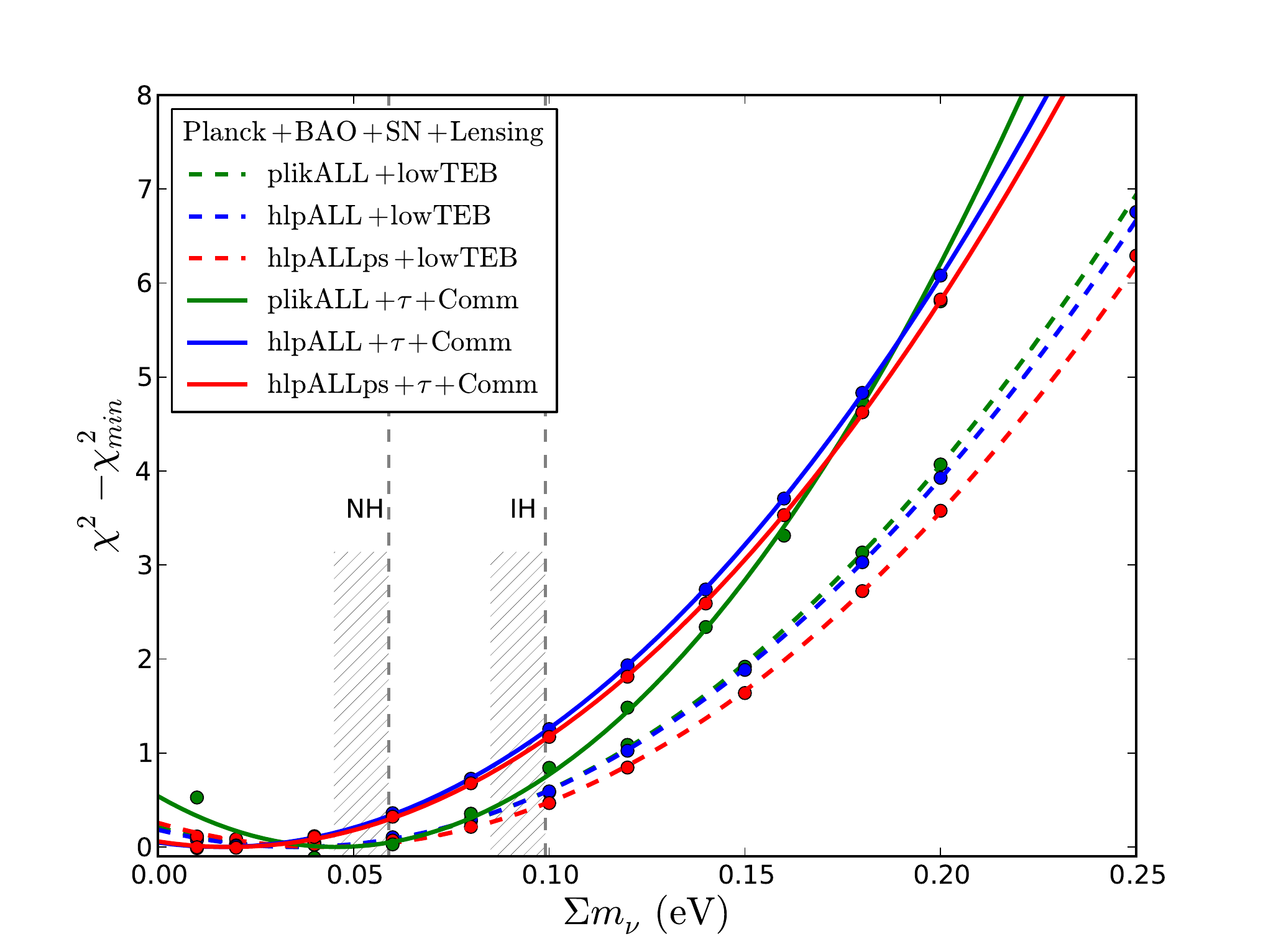}
\caption{\label{fig:cmp_all_lens} \mnu\ profile  likelihoods obtained when combining either \plikALL, \hlpALL,
and \hlpALLps, temperature+polarisation likelihoods,  with the CMB lensing likelihood, 
\BAO\ and \SNplot\ for \lowTEB\ and for the combination of an auxiliary constraint on $\tau$+\Commander.
We also materialise the minimal neutrino masses for the normal and
inverted hierarchy inferred from neutrino oscillation measurements.}
\end{figure}

As done in Sect.~\ref{sec:lens1}, we now add to the 
data combination, the lensing \planck\ likelihood (see Sect.~\ref{lensing}).
The corresponding profile likelihoods are  shown in Fig.~\ref{fig:cmp_all_lens},
and the results are given in Table \ref{tab:limits_baosnALL_lens} for \MnutroisnulambdaCDM\ (i.e. with $\alens=1$). 
To compare with Table \ref{tab:limits_baosnALL}, the \mnu\ limits are higher  
when \lowTEB\ is used at low-$\ell$, 
but more robust with respect to the \alens\ issue thanks to the use of the
lensing data. 
For the ALL case, in the \troisnulambdaCDM+\alens\ model we
end up with a value of \alens\ compatible with one and very comparable
with those of Table~\ref{tab:limits_baosnTT_lens}.  
The limits on \mnu\ are therefore not artificially lowered 
by an overly high \alens\ value. Even though we end up with upper limits
that are pushed toward higher bounds if compared to those obtained without the lensing
data, we insist on the fact that this data combination is compatible
with the \lambdaCDM\ model.

\begin{table}
\centering          
\begin{tabular}{l c c } 
\hline\hline       
\hbox{\planck ALL} &  +\lowTEB & +$\taureio$\\
\hbox{+\SNplot+\BAO+lensing} & & +\Commander  \\
\hline                    
$\hbox{\hlpALL }$& 0.20  & 0.16 \cr 
$\hbox{\hlpALLps}$& 0.21 & 0.17 \cr 
$\hbox{\plikALL} $& 0.19  & 0.17 \cr 
\hline                  
\end{tabular}
\caption{95$\%$ CL upper limits on \mnu\ in  \MnutroisnulambdaCDM\
obtained when combining \plikALL, \hlpALL\ or \hlpALLps\  with \SNplot+\BAO+lensing,
using \lowTEB\ for the low-$\ell$ (second
column) and for the combination of an auxiliary constraint on $\taureio$+\Commander (third column) .
ALL refers to the combination TT+TE+EE. 
\label{tab:limits_baosnALL_lens} }
\end{table}

When making use of the latest \taureio\ measurement, we 
almost recover the results of Table \ref{tab:limits_baosnALL}.
We use the differences between the upper limits obtained with the three
\planck\ likelihoods of Table \ref{tab:limits_baosnALL_lens} (last column) to 
estimate a systematic error coming from the foreground
modelling of 0.01\eV.

Table~\ref{tab:fit-data} provides the $\chi^2=-2\log\cal{L}$ values as a function
of \mnu, where the likelihood ($\cal{L}$) has been profiled out
over the nuisance and cosmological parameters. It corresponds to
the combination of \hlpALLps+\BAO+\SN+lensing, using the auxiliary constraint
on \taureio\ combined with \Commander\ at low-$\ell$. This dataset
is chosen for the final limits derivation since it corresponds to 
the most up-to-date results on \taureio. Table~\ref{tab:fit-data}
can be used for neutrino global fits. 

\begin{table}
\centering          
\begin{tabular}{c c } 
\hline\hline       
\mnu (\eV) &  $\chi^2=-2\log\cal{L}$  \\
\hline
0.01 & 28613.76 \\
0.02 & 28613.75 \\
0.04 & 28613.86 \\
0.06 & 28614.08 \\
0.08 & 28614.44 \\
0.10 & 28614.93 \\
0.12 & 28615.57 \\
0.14 & 28616.35 \\
0.16 & 28617.29 \\
0.18 & 28618.38 \\
0.20 & 28619.58 \\
0.25 & 28623.26 \\
0.30 & 28627.86 \\
0.35 & 28633.30 \\
0.40 & 28639.60 \\
\hline                    
\end{tabular}
\caption{
\label{tab:fit-data} Values of the $\chi^2=-2\log\cal{L}$ profiled out over all the other (cosmological
and nuisance) parameters
as a function of \mnu\ for the \hlpALLps+\BAO+\SN+lensing combination, using the auxiliary constraint
on \taureio\ combined with \Commander\ at low $\ell$. They correspond to the red dots/plain line
of Fig.~\ref{fig:cmp_all_lens}. }
\end{table}

\subsection{Cosmological parameters: \texorpdfstring{\lambdaCDM}{LambdaCDM} versus \texorpdfstring{\MnulambdaCDM}{MnuLambdaCDM}}
\label{sec:lcdm_parms}

We compare  the  \lambdaCDM\ cosmological 
parameters and their error bars derived
with the profile likelihood method using various combinations of CMB temperature+polarisation high-$\ell$ and low-$\ell$  likelihoods,
with the  CMB lensing likelihood from \planck, \BAO\ and \SN\ datasets.

\begin{figure*}[htb]
        \begin{tabular}{ccc}
        \includegraphics[width=.29\textwidth]{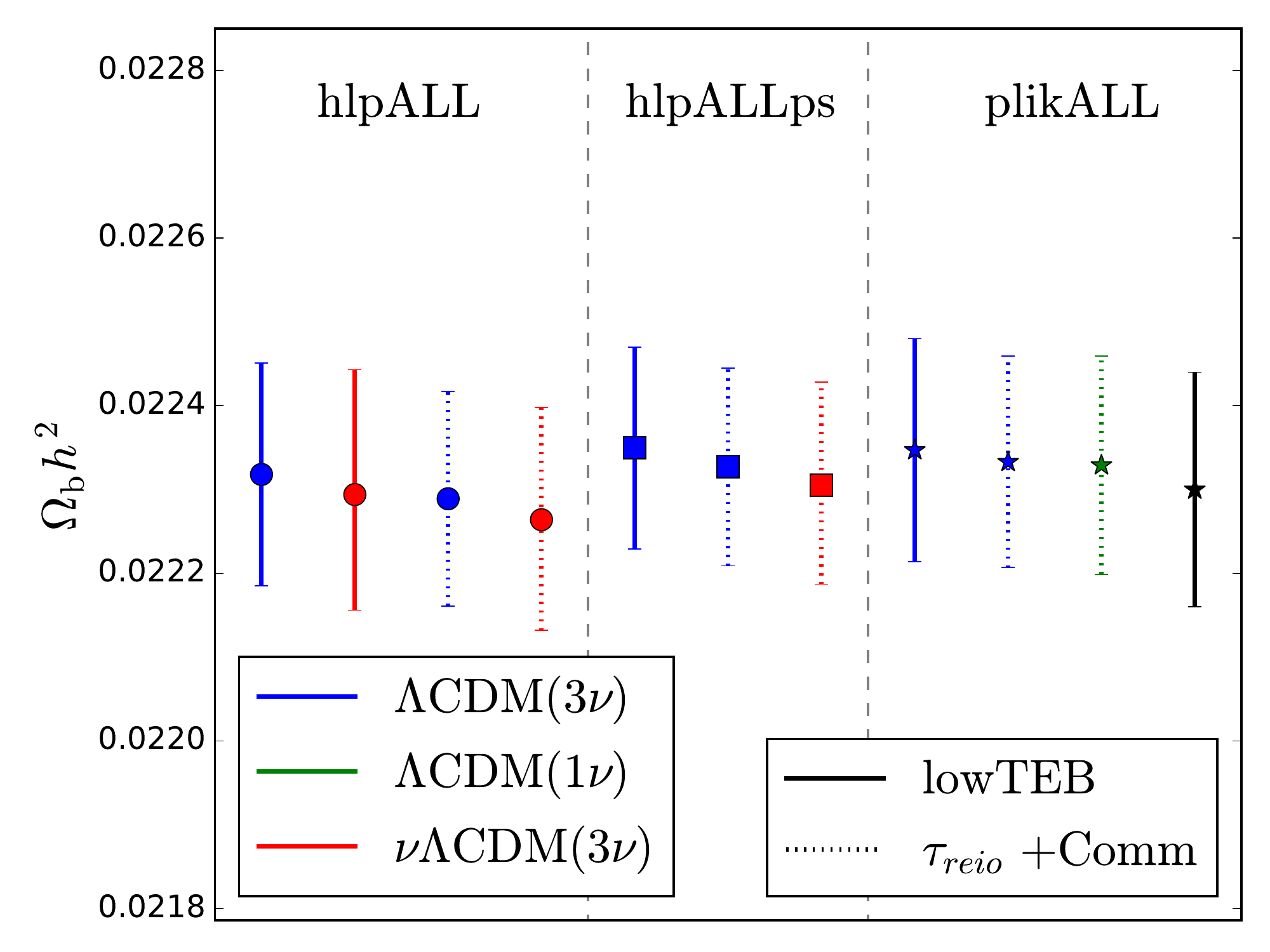}
        & 
        \includegraphics[width=.29\textwidth]{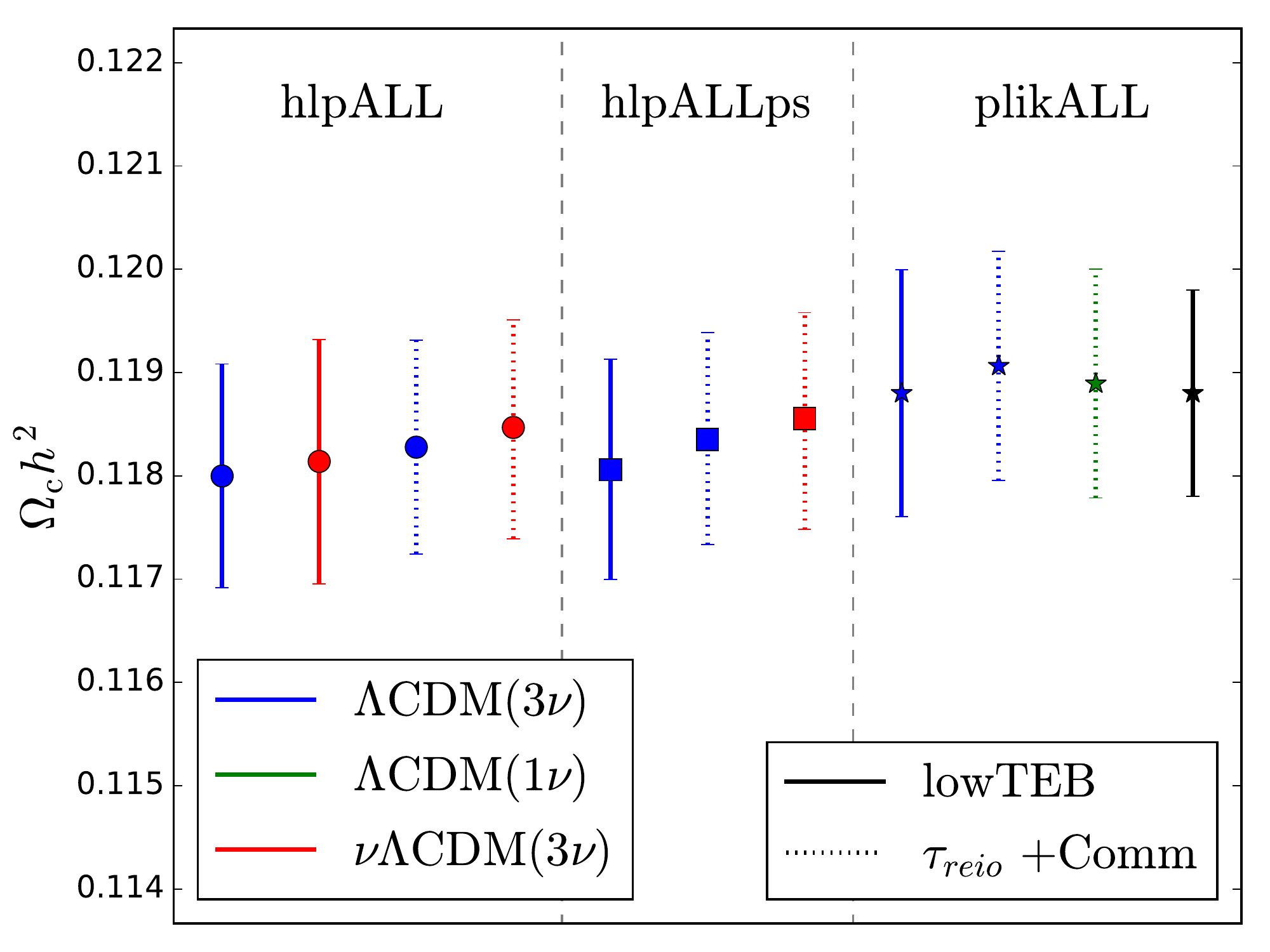}
        & 
        \includegraphics[width=.29\textwidth]{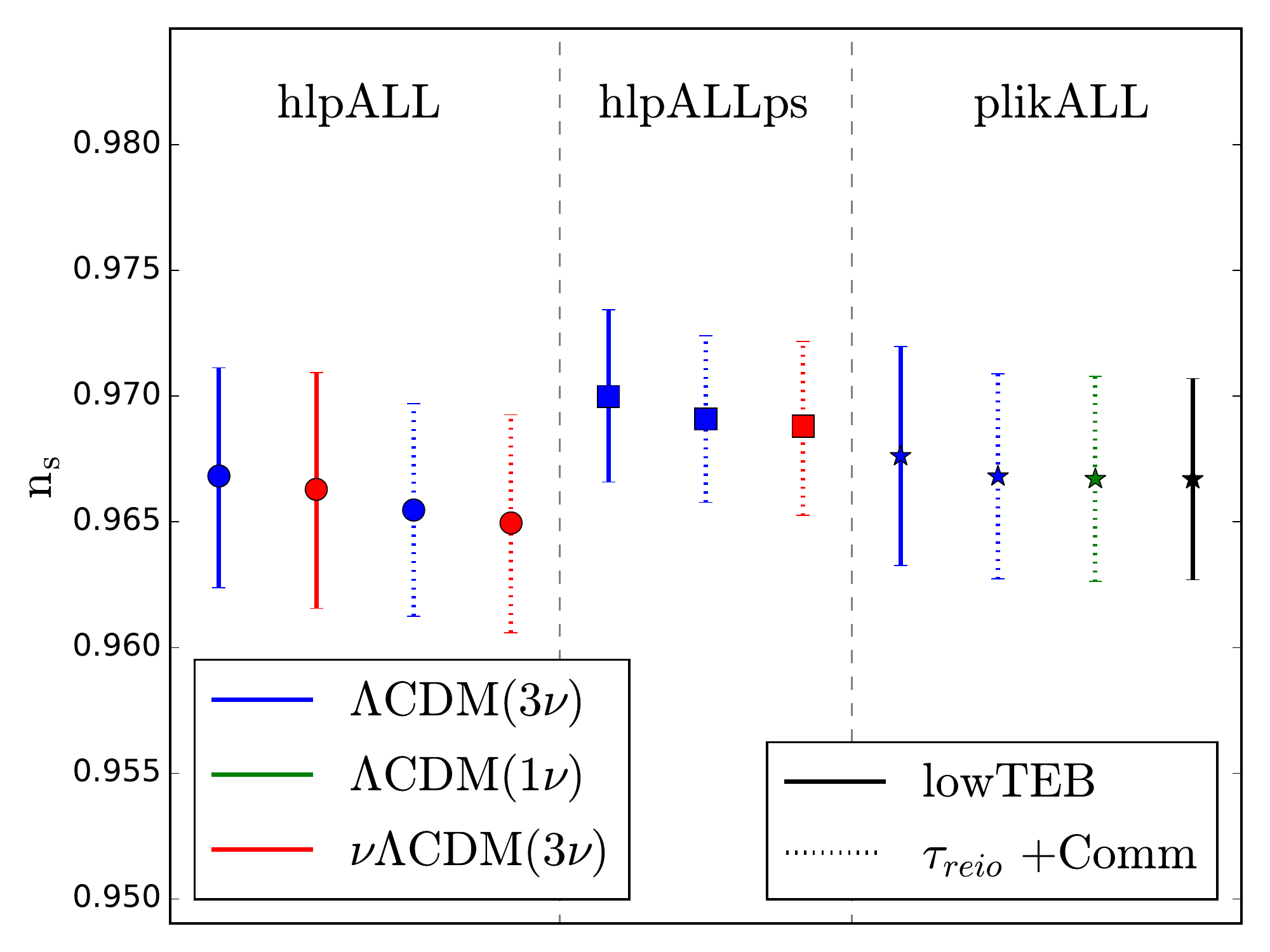}
        \\
        \includegraphics[width=.29\textwidth]{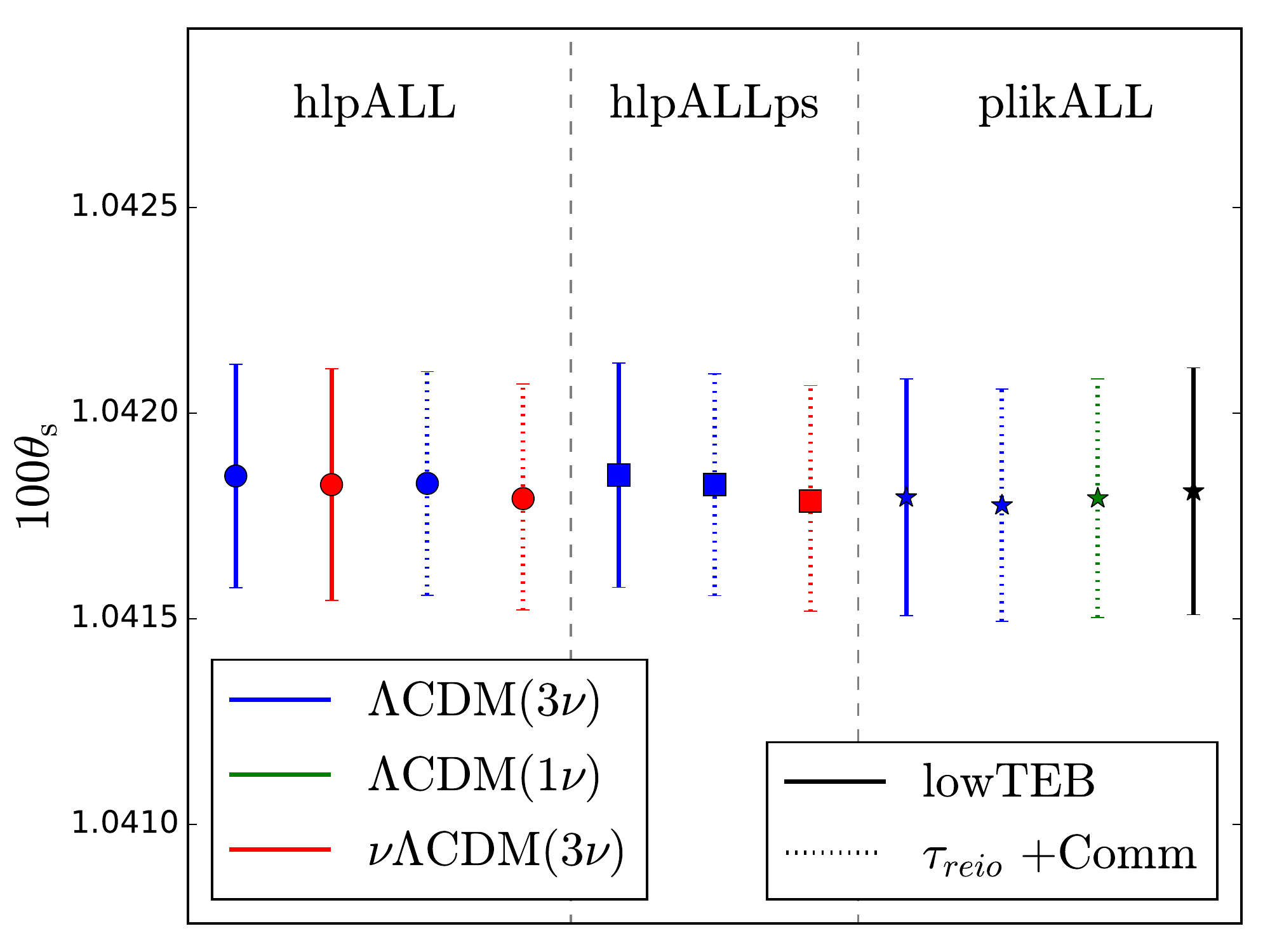}
        & 
        \includegraphics[width=.29\textwidth]{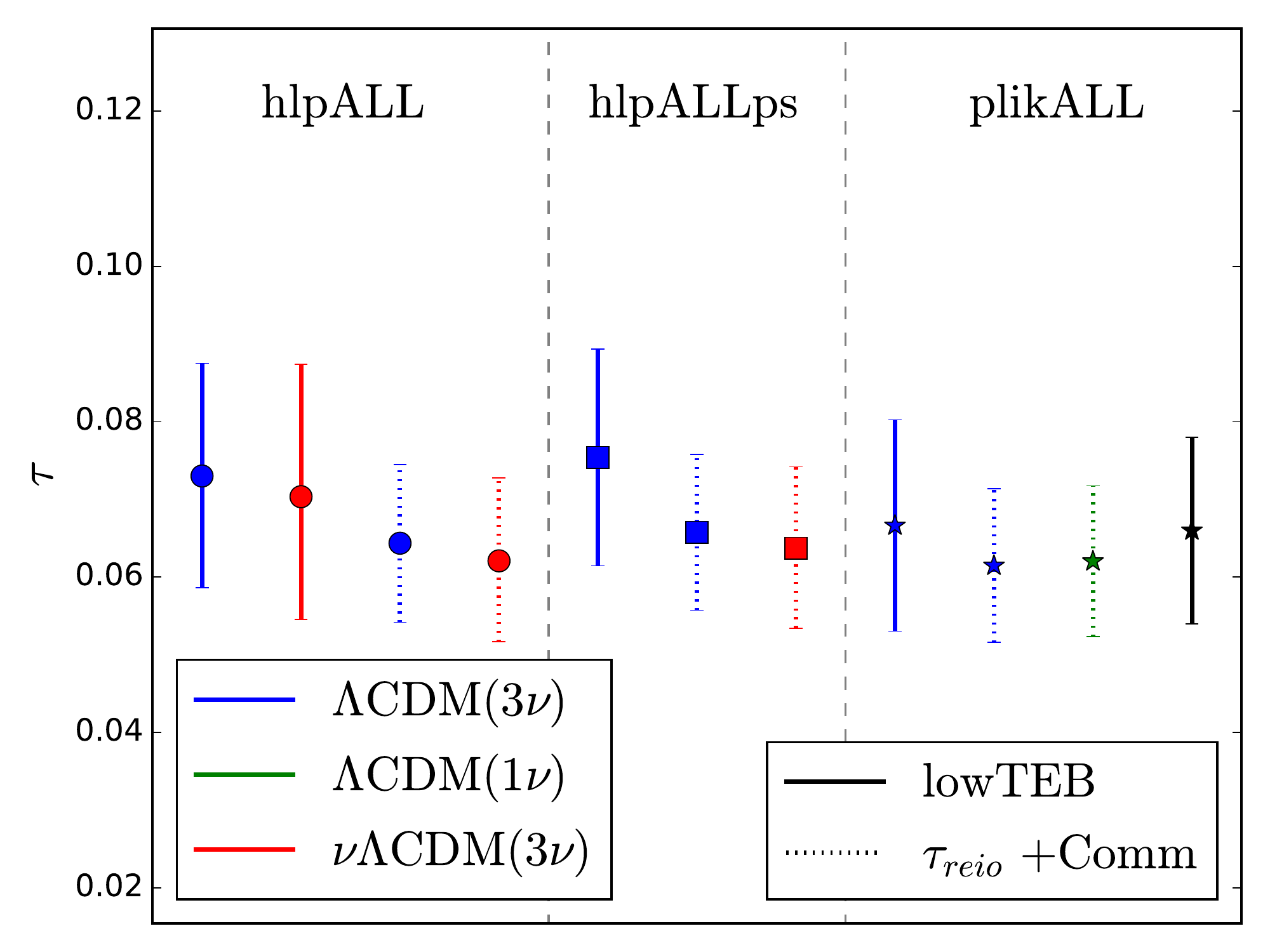}
        & 
        \includegraphics[width=.29\textwidth]{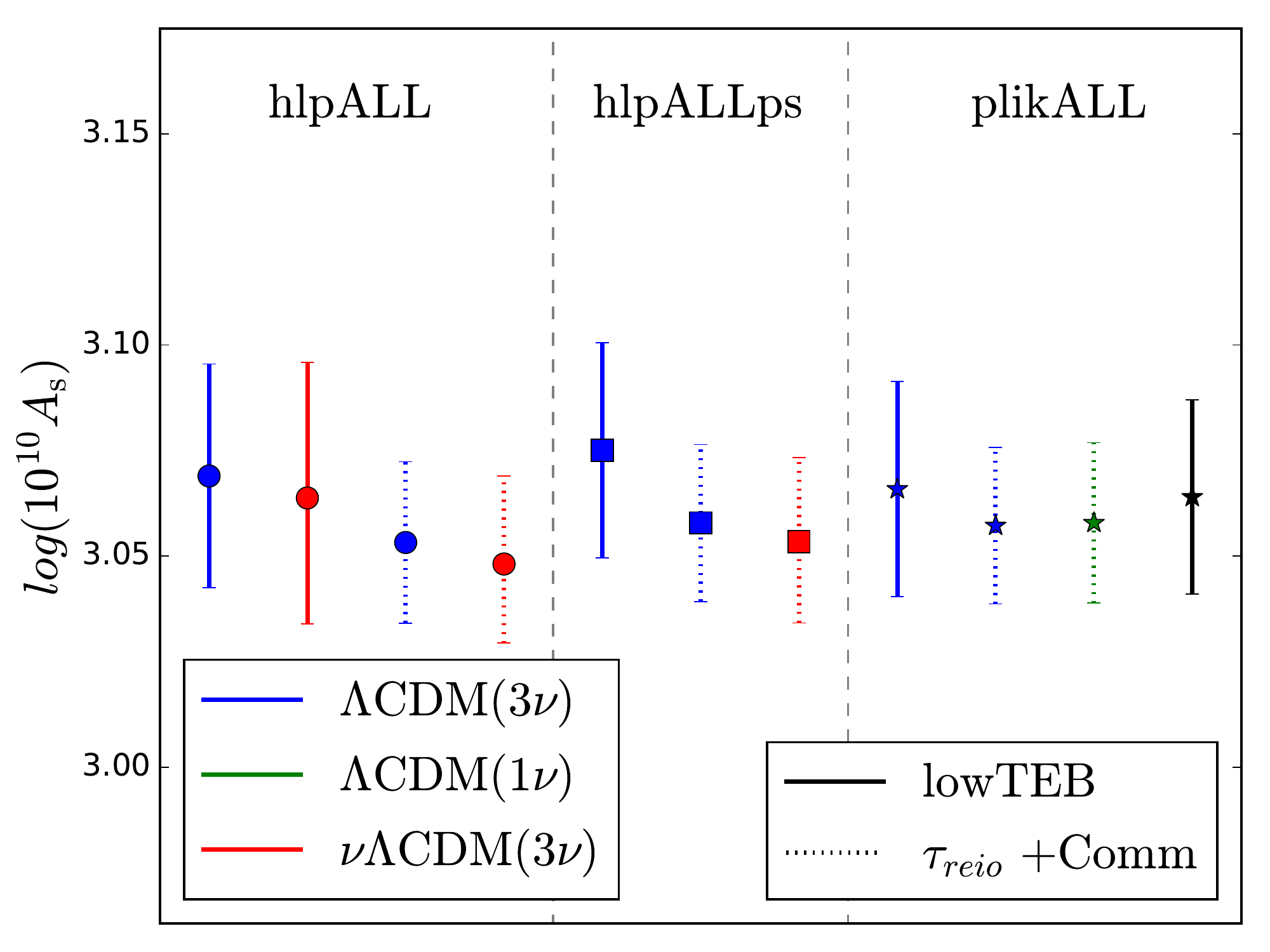}
        \end{tabular}
        \caption{Comparison of various estimations of cosmological parameters, together with their $68\%$ CL, in the  \troisnulambdaCDM, \MnutroisnulambdaCDM\ and \baselambdaCDM\ models, from the combination of: the high-$\ell$ \planck\ (\hlpALL, \hlpALLps\ or \plikALL\ separated by the vertical dashed lines); the \lowTEB\ likelihood or a \taureio\ auxiliary constraint;  \Commander; the CMB lensing from \planck; \BAO; and \SN. Those results, derived from profile likelihood analyses, are compared (last point in black)  to the \planck\ 2015 results with a similar data combination (last column of Table 4, in \cite{planck2014-a15}).}
%Those results, derived from profile likelihood analyses, are compared to the \planck\ 2015 results from \citet{planck2014-a15} (last point in black).}
        \label{fig:parm_results}
\end{figure*}

More precisely, this comparison is done: 
\begin{enumerate}
\item when \mnu\ is, or not, a free parameter,
\item using different foreground-modelling choices (via the different high-$\ell$ likelihoods),
\item switching from the publicly available \lowTEB\ low-$\ell$ likelihood 
to the combination of an auxiliary constraint on \taureio\ with \Commander, to size the impact of a
tighter constraint on \taureio,
\item between the neutrino mass settings of the \baselambdaCDM\ and \troisnulambdaCDM\ models.
\end{enumerate} 
These results are summarised in Fig.~\ref{fig:parm_results}. 
They are very similar to the \planck\ 2015 results \citep{planck2014-a15} even though we are using here a new version of the BAO data (DR12).
As stated in Sect.~\ref{sec::prof}, we have checked that they do not depend on the
chosen statistical approach (Bayesian or Frequentist), either for the \lambdaCDM\ or for the \MnulambdaCDM\ model.

The values and uncertainties of the cosmological parameters 
in  the  \MnutroisnulambdaCDM\ model (in red) are similar  to those 
obtained in   \troisnulambdaCDM\  (in blue), but are marginally shifted and with slightly larger $68\%$ CL uncertainties. This is 
true with \lowTEB\  (as seen from the \hlpALL\ results, circles) as well as with 
an auxiliary constraint on \taureio\ with \Commander\  for both \hlpALL\ and \hlpALLps\ (shown with squares). 
The increase of the uncertainties is related to the addition of \mnu\ in the fit.
The small shifts of the mean values are nearly the same for all the tested cases.
This could be the result of a 
best fit value of \mnu\ slightly different from 0.06~\eV assumed in the
\troisnulambdaCDM\ model.

Switching from \lowTEB\ (plain line on Fig.~\ref{fig:parm_results})  to an auxiliary constraint on \taureio\ + \Commander\ (dotted lines) 
at low-$\ell$ changes the results on \taureio\ and \As\ and reduces their uncertainties, as expected.
We observe small shifts on other parameters ($\omega_{\rm b}$, $\omega_{\rm cdm}$, \ns),  
consistently for all three high-$\ell$ likelihoods, when fitting for \mnu. They result from 
intrinsic correlations between (\taureio, \As) and the other cosmological parameters.

In the six-parameter   \troisnulambdaCDM\ case, \hlpALL\ and \hlpALLps\ give very similar results, but for a small difference on $n_s$.
This is related to the more constraining point source model \citep[we refer to the discussion in][]{Couchot:2016vaq}. 
The comparison, illustrated in Fig~\ref{fig:parm_results},
shows the robustness of the cosmological parameters estimation with respect to the choice
of the CMB (high-$\ell$ and low-$\ell$) likelihoods. The residual (small) differences between them illustrate the
remaining systematic uncertainties. For example, the differences between
\plik\ and \hlp\,  could be linked to 
the different choices made for masks, $\ell$ ranges and foreground templates used in both cases.

Finally, the values and uncertainties of the cosmological parameters fitted in the \troisnulambdaCDM\ and \baselambdaCDM, with \plikALL, 
are very close to each other. This  shows that the mass repartition has almost no effect on \lambdaCDM\ parameters
when \mnu\ is fixed to 0.06~\eV.

\section{Conclusions}

We have addressed the question of the propagation of foreground systematics
on the determination of the sum of the neutrino masses
through an extensive comparison of results derived from the combination
of cosmological data including \planck\ CMB likelihoods with
different foreground modelisations.

For this comparison we have worked within the \MnutroisnulambdaCDM\ model
assuming three mass-degenerate neutrinos, motivated by oscillations results.
We have justified this approximation, showing that it leads to the
same results as those obtained when considering normal or inverted hierarchy.

We have shown that the details of the foreground residuals modelling play
a non-negligible role in the \mnu\ determination, and affect the results in two different ways.
Firstly, they are unveiled by different \alens\ values for the various likelihoods,
up to $2\sigma$ away from \lambdaCDM.
This impacts
the \mnu\ limit: The higher the \alens\ value favoured by the data, the lower the upper bound
on \mnu. 
For this reason we have added the CMB lensing in the
combination of data, 
providing a way to derive a limit with an \alens\ value fully
compatible with the \lambdaCDM\ model.
Secondly, 
it introduces
a spread of the profile likelihoods, resulting in various limits on
\mnu, from which a systematic uncertainty was derived.
We have compared CMB temperature and polarisation results, 
as well as their combination, and showed that the
results are very consistent between themselves.

We have also discussed the impact of the low-$\ell$ likelihoods.
We have shown, through the use of an auxiliary constraint
on $\taureio$ (derived from the
latest \planck\ reionisation results) 
combined with \Commander, that a better determination of the uncertainty on \taureio\
led to a reduction of the upper limit on \mnu, of the order of a few $10^{-2}$~\eV\
with respect to the \lowTEB\ case.

We have also addressed the question of the neutrino hierarchy.
We have shown that the profile likelihoods are identical in the normal and
inverted hierarchies, proving that the current data are not sensitive to the
details of the mass repartition. Still, cosmological data could rule out
the inverted hierarchy if they lead to a low-enough \mnu\ limit. 
However, today, the \mnu\ upper bound
is still too high to get to this conclusion.

Combining the latest results from CMB anisotropies with \planck\ (both
in temperature and polarisation, and including the last measurement
of \taureio),  
with \BAO, \SN, and the CMB lensing,
we end up with:
$$\mnu<\ 0.17\ [ \hbox{incl.}\ 0.01\ \hbox{(foreground syst.)} ] \hbox{\ eV\ at\ }95\%\hbox{\ CL}\ .$$
The values of the $\chi^2$ of the profile likelihoods are also given for further use in
neutrinos global fits.
For the first time, all the following effects have been taken into account:
\begin{itemize}
\item Systematic variations related to foreground modelling error,
\item a value of \alens\ compatible with expectations,
\item a lower value for \taureio\ compatible with the latest measurements from \planck,
\item the new version of the \BAO\ data (DR12),
\end{itemize}
making our final \mnu\ limit a robust result.
%The first two effects tend to higher up the limit on \mnu, while the two last result in a lower upper limit.
For all these reasons, we think that this is the lowest upper limit we can obtain today using cosmological data.

%\comment{For this result, the fact that the data tend to favor \alens\ greater than one is taken care of, 
%as well as the impact of the choice of the foreground modelling errors. Both effects tend to 
%higher up the constraint on \mnu. Since, in addition, we
%have considered new BAO data and an update of the \taureio\ measurement which result in a lower
%upper limit, we end up with a result very similar to~\cite{planck2014-a15}: this is just a coincidence.}

As far as cosmology is concerned, the uncertainty on the neutrino mass will be improved in the future:
It could be reduced by a factor $\simeq 5$ if one refers, for instance, to the forecasts on the combination
of next-generation `Stage 4' B-mode CMB experiments 
with BAO and clustering measurements from DESI~\citep{Audren:2012vy,Font-Ribera:2013rwa,Allison:2015qca,Abazajian:2016yjj,Archidiacono:2016lnv}.
Nevertheless, the proper propagation of systematics, in particular coming from the modelling of
foregrounds, is a more important topic than 
ever in today's cosmology.

%%%%%%%%%%%%%%%%%%%%%%%%%%%%%%%%%%%%%%%%%%%%%%%%%%%%%%%%%%%%%%%%%%%%%%%%%%
%ACKNOWLEDGEMENT
%%%%%%%%%%%%%%%%%%%%%%%%%%%%%%%%%%%%%%%%%%%%%%%%%%%%%%%%%%%%%%%%%%%%%%%%%%
\begin{acknowledgement}
We gratefully acknowledge the IN2P3 Computer Center (http://cc.in2p3.fr) for providing the computing resources and services needed to this work.
\end{acknowledgement}

%\newpage
%%%%%%%%%%%%%%%%%%%%%%%%%%%%%%%%%%%%%%%%%%%%%%%%%%%%%%%%%%%%%%%%%%%%%%%%%%
%BIBLIOGRAPHY
%%%%%%%%%%%%%%%%%%%%%%%%%%%%%%%%%%%%%%%%%%%%%%%%%%%%%%%%%%%%%%%%%%%%%%%%%%
\bibliographystyle{aat} % style aa.bst
\bibliography{refs}

%%%%%%%%%%%%%%%%%%%%%%%%%%%%%%%%%%%%%%%%%%%%%%%%%%%%%%%%%%%%%%%%%%%%%%%%%%
%ANNEXES
%%%%%%%%%%%%%%%%%%%%%%%%%%%%%%%%%%%%%%%%%%%%%%%%%%%%%%%%%%%%%%%%%%%%%%%%%%

\end{document}